\documentclass[manuscript]{aastex}
\usepackage{amsmath}
\usepackage{arydshln}
\usepackage{lscape}
\usepackage{comment}

\newcommand{\cch}{$N$=3--2, $J$=7/2--5/2, $F$=4--3 and 3--2}
\newcommand{\cs}{$J$=$5$--$4$}
\newcommand{\so}{$J_N$=$6_7$--$5_6$}
\newcommand{\hnco}{$12_{0, 12}$--$11_{0, 11}$}
\newcommand{\fa}{$12_{0, 12}$--$11_{0, 11}$}
\newcommand{\FA}{NH$_2$CHO}
\newcommand{\sio}{$J$=6--5}
\newcommand{\mf}{$20_{5, 16, 0}$--$19_{5, 15, 0}$}
\newcommand{\MF}{HCOOCH$_3$}
\newcommand{\hcooh}{$12_{0, 12}$--$11_{0, 11}$}
\newcommand{\dee}{$13_{5, 8, 1}$--$13_{4, 9, 1}$}
\newcommand{\DEE}{(CH$_3$)$_2$O}
\newcommand{\aal}{$14_{1, 14}$--$13_{1, 13}$; A}
\newcommand{\AAL}{CH$_3$CHO}
\newcommand{\FAD}{H$_2$CO}
\newcommand{\HHCS}{H$_2$CS}

\newcommand{\MET}{CH$_4$}
\newcommand{\MN}{CH$_3$OH}

\newcommand{\Msun}{$M_\odot$}
\newcommand{\Lsun}{$L_\odot$}
\newcommand{\rCB}{$r_{\rm CB}$}

\newcommand{\vlsr}{$v_{\rm LSR}$}

\newcommand{\kmps}{km s$^{-1}$}
\newcommand{\inv}{$^{-1}$}

\shorttitle{L483: hot corino activity}
\shortauthors{Oya et al.}

\title{L483: Warm Carbon-Chain Chemistry Source Harboring Hot Corino Activity}
\author{Yoko Oya\altaffilmark{1}, Nami Sakai\altaffilmark{2}, Yoshimasa Watanabe\altaffilmark{1}, 
Aya E. Higuchi\altaffilmark{2}, Tomoya Hirota\altaffilmark{3}, \\ 
Ana L{\'o}pez-Sepulcre\altaffilmark{1, 4}, Takeshi Sakai\altaffilmark{5}, Yuri Aikawa\altaffilmark{6}, 
Cecilia Ceccarelli\altaffilmark{7, 8}, \\Bertrand Lefloch\altaffilmark{7, 8}, 
Emmanuel Caux\altaffilmark{9, 10}, Charlotte Vastel\altaffilmark{9, 10}, Claudine Kahane\altaffilmark{7, 8}, 
\\and Satoshi Yamamoto\altaffilmark{1}} 

\email{oya@taurus.phys.s.u-tokyo.ac.jp}
\altaffiltext{1}{Department of Physics, The University of Tokyo, 7-3-1, Hongo, Bunkyo-ku, Tokyo 113-0033, Japan}
\altaffiltext{2}{The Institute of Physical and Chemical Research (RIKEN), Wako, Saitama 351-0198, Japan}
\altaffiltext{3}{National Astronomical Observatory of Japan, Osawa, Mitaka, Tokyo 181-8588, Japan}
\altaffiltext{4}{Institut de Radioastronomie Millim\'etrique (IRAM), 38406, Saint Martin d'H\'eres, France}
\altaffiltext{5}{Department of Communication Engineering and Informatics, Graduate School of Informatics and Engineering, The University of Electro-Communications, Chofugaoka, Chofu, Tokyo 182-8585, Japan}
\altaffiltext{6}{Center for Computational Science, University of Tsukuba, Tsukuba, Ibaraki 305-8577, Japan}
\altaffiltext{7}{Universite Grenoble Alpes, IPAG, F-38000 Grenoble, France}
\altaffiltext{8}{CNRS, IPAG, F-38000 Grenoble, France}
\altaffiltext{9}{Universite de Toulouse, UPS-OMP, F-31028 Toulouse Cedex 4, France}
\altaffiltext{10}{CNRS, IRAP, 9 Av. Colonel Roche, BP 44346, F-31028 Toulouse Cedex 4, France}

\begin{abstract}
The Class 0 protostar, L483, has been observed in various molecular lines 
in the 1.2 mm band at a sub-arcsecond resolution with ALMA. 
An infalling-rotating envelope is traced by the CS line, 
while a very compact component with a broad velocity width is 
observed for the CS, SO, HNCO, \FA, and \MF\ lines. 
Although this source is regarded as the warm carbon-chain chemistry (WCCC) candidate source at a 1000 au scale, 
complex organic molecules characteristic of hot corinos such as \FA\ and \MF\ are detected in the vicinity of the protostar. 
Thus, both hot corino chemistry and WCCC are seen in L483. 
Although such a mixed chemical character source has been recognized as an intermediate source in previous single-dish observations, 
we here report the first spatially-resolved detection. 
A kinematic structure of the infalling-rotating envelope is roughly explained by a simple ballistic model 
with the protostellar mass of 0.1--0.2 \Msun\ and the radius of the centrifugal barrier (a half of the centrifugal radius) of 30--200 au, 
assuming the inclination angle of 80\degr\ (0\degr\ for a face-on). 
The broad line emission observed in the above molecules most likely comes from the disk component inside the centrifugal barrier. 
Thus, a drastic chemical change is seen around the centrifugal barrier. 
\end{abstract}


\begin{document}
\section{Introduction} \label{sec:intro}
It is now recognized that low-mass star-forming regions show a significant chemical diversity, 
even if their evolutionary stages are almost the same (Class 0/Class I). 
Two distinct families so far identified are hot corino chemistry characterized by saturated complex organic molecules (COMs) 
(e.g. Cazaux et al. 2003; J\o rgensen et al. 2005), 
and warm carbon-chain chemistry (WCCC) characterized by unsaturated species (carbon-chain molecules) (e.g. Sakai et al. 2008a, 2009a). 
A prototypical source for hot corino chemistry is IRAS 16293--2422 in Ophiuchus, while that for WCCC is L1527 in Taurus. 
Exclusive chemical compositions can be seen in these prototypical sources: 
carbon-chain molecules are deficient in the hot corino IRAS 16293--2422, 
while COMs are not detected in the WCCC source L1527 (Sakai et al. 2008b; Sakai \& Yamamoto 2013). 
In addition, the existence of intermediate sources is also suggested 
by single-dish observations (e.g. Sakai et al. 2009a). 
Both origin and evolution of the chemical diversity are important targets for astrochemistry. 

Such diversity seems to originate from different chemical composition of the ice mantle at the onset of star formation (Sakai \& Yamamoto 2013): 
WCCC sources are rich in \MET, while hot corinos are rich in saturated organic molecules. 
It is proposed that a duration time of the starless core after shielding the interstellar UV radiation could cause such a difference of the ice composition 
(Sakai et al. 2009a; Sakai \& Yamamoto 2013). 
This mechanism 
is supported by a low deuterium fractionation in the WCCC source L1527 (Sakai et al. 2009b). 
On the other hand, it is also suggested that the environmental effects such as local variation of the UV radiation field 
and the effects by nearby star-formation activities could be responsible for the chemical diversity 
(Lindberg et al. 2015; Spezzano et al. 2016). 
In any case, 
sources whose chemical compositions are intermediate between those of hot corinos and WCCC sources (or a mixture of the two) are naturally expected. 
Recent infrared observations indeed indicate the coexistence of \MET\ and \MN\ ices (Graninger et al. 2016). 
This result implies that the intermediate sources could be common occurrence. 
It is thus important to characterize such intermediate sources for understanding the chemical diversity and its origin. 

On the other hand, the evolution of the chemical diversity to the protostellar disk has 
recently been studied toward both hot corinos and WCCC sources at a high spatial resolution with ALMA 
(Sakai et al. 2014a, 2014b, 2016; Oya et al. 2014, 2015, 2016; Imai et al. 2016). 
The observations revealed that these protostellar cores have different chemical compositions even at a spatial scale of a few tens of au around a protostar, 
where a disk structure is being formed. 
More importantly, 
the chemical compositions drastically change near the protostar. 
In the WCCC sources L1527 (Sakai et al. 2014a, 2014b) and TMC--1A (Sakai et al. 2016), 
carbon-chain molecules as well as CS exist in an envelope rotating and infalling toward its centrifugal barrier at radii of 100 au and 50 au, respectively, 
while SO is concentrated around the centrifugal barrier. 
On the other hand, \FAD\ is present ubiquitously in the envelope, the centrifugal barrier, and the disk inside the barrier. 
In the hot corino source IRAS 16293--2422 A, 
C$^{34}$S and OCS exist in an infalling-rotating envelope, 
COMs mainly around its centrifugal barrier, 
and \HHCS\ in both the envelope and disk components (Oya et al. 2016). 

Chemical composition in the vicinity of the protostar is of fundamental importance, 
because it will define interstellar chemical heritage to protoplanetary disks. 
However, a few sources, 
including two WCCC sources (L1527 and TMC--1A; Sakai et al. 2014a, 2014b, 2016; Oya et al. 2015) and 
three hot corinos (IRAS 2A, IRAS 16293--2422 A, and B335; Maury et al. 2014; Oya et al. 2016; Imai et al. 2016), 
have been studied so far for chemical characterization at a 100 au scale. 
Hence, observations of other protostellar sources including the sources with intermediate chemical compositions are 
still awaited. 
In this study, we focus on the well-studied Class 0 protostar in L483. 

The L483 dark cloud is located in the Aquila Rift ($d$ = 200 pc; J\o rgensen et al. 2002; Rice et al. 2006); 
which harbors the Class 0 protostar IRAS 18148--0440 
(Fuller et al. 1995; Chapman et al. 2013). 
Its bolometric luminosity is 13 $L_\odot$ (Shirley et al. 2000). 
We here adopt the systemic velocity of 5.5 km s\inv\ for this source based on previous single-dish observations (Hirota et al. 2009). 
In this source, 
the C$_4$H abundance is relatively high, 
and it is regarded as a possible candidate for the WCCC source (Sakai et al. 2009a; Hirota et al. 2010; Sakai \& Yamamoto 2013). 
Recent detection of the novel carbon-chain radical HCCO further supports the carbon-chain-molecule rich nature of this source (Ag\'undez et al. 2015). 
The outflow of L483 has extensively been studied 
(Fuller et al. 1995; Hatchell et al. 1999; Park et al. 2000; Tafalla et al. 2000; J\o rgensen 2004; Takakuwa et al. 2007; Leung et al. 2016). 
It is extended along the east-west axis, 
where the eastern and western components are red-shifted and blue-shifted, respectively. 
The position angle of the outflow axis is reported to be 95\degr\ by Park et al. (2000) based on the  HCO$^+$ ($J$=1--0) observation 
and 105\degr\ by Chapman et al. (2013) based on the shocked H$_2$ emission reported by Fuller et al. (1995). 
The inclination angle of the outflow axis is reported to be $\sim$50\degr\ with respect to the plane of the sky (0\degr\ for a face-on configuration) (Fuller et al. 1995). 
Along the line perpendicular to the outflow, the northern part is blue-shifted, 
while the southern part is red-shifted, 
according to the CS ($J$=2--1, 7--6) and HCN ($J$=4--3) observations (J\o rgensen 2004; Takakuwa et al. 2007). 
This suggests a rotating motion of the envelope. 
Chapman et al. (2013) reported the position angle of a pseudo-disk to be 36\degr\ based on their {\it Spitzer} 4.5 $\mu$m observation. 

In these previous studies, the disk/envelope system is not well resolved, 
and little is known about a chemical composition in the closest vicinity of the protostar. 
In the present study, we investigate the physical and chemical structures around the protostar at a 100 au scale with ALMA. 

\section{Observation} \label{sec:observation}
The ALMA observation of L483 was carried out in the Cycle 2 operation on 12 June 2014. 
Spectral lines of CCH, CS, SO, HNCO, t-HCOOH, \AAL, \FA, \MF, \DEE, and SiO were observed 
with the Band 6 receiver in the frequency range from 244 to 264 GHz (Table \ref{tb:lines}).  
Thirty-four antennas were used in the observation, 
where the baseline length ranged from 18.5 to 644 m. 
The field center of the observations was ($\alpha_{2000}$, $\delta_{2000}$) = ($18^{\rm h} 17^{\rm m} 29\fs910$, $-04\degr 39\arcmin 39\farcs60$). 
The primary beam size (FWHM) is 23\farcs03. 
The total on-source time was 25.85 minutes, 
where a typical system temperature was 60--100 K. 
A backend correlator was tuned to a resolution of 61.030 kHz 
and a bandwidth of each chunk of 58.5892 MHz. 
The resolution corresponds to the velocity resolution of 0.073 km s$^{-1}$ at 250 GHz. 
J1733-1304 was used for the bandpass calibration and for the phase calibration every 7 minutes. 
An absolute flux density scale was derived from Titan. 
The data calibration was performed in the antenna-based manner, and uncertainties are less than 9\%. 

Images were obtained by using the CLEAN algorithm, 
where the Brigg's weighting with the robustness parameter of 0.5 was employed. 
We applied self-calibration for better imaging. 
A continuum image was prepared by averaging line-free channels, 
and the line maps were obtained after subtracting the continuum component directly from the visibilities. 
Synthesized-beam sizes for the spectral lines are listed in Table \ref{tb:lines}. 
An rms noise level for the continuum is 0.13 mJy beam$^{-1}$, 
while those for 
CCH, CS, SO, HNCO, \FA, \MF, and SiO maps are derived from the nearby line free channels to be 
8.2, 7.6, 6.1, 6.5, 4.4, 5.8, and 3.5 mJy beam$^{-1}$, respectively, for the channel width of 61.030 kHz. 

\section{Distribution} \label{sec:distribution}
\begin{figure}
	\begin{center}
	\epsscale{1.0}
	\plotone{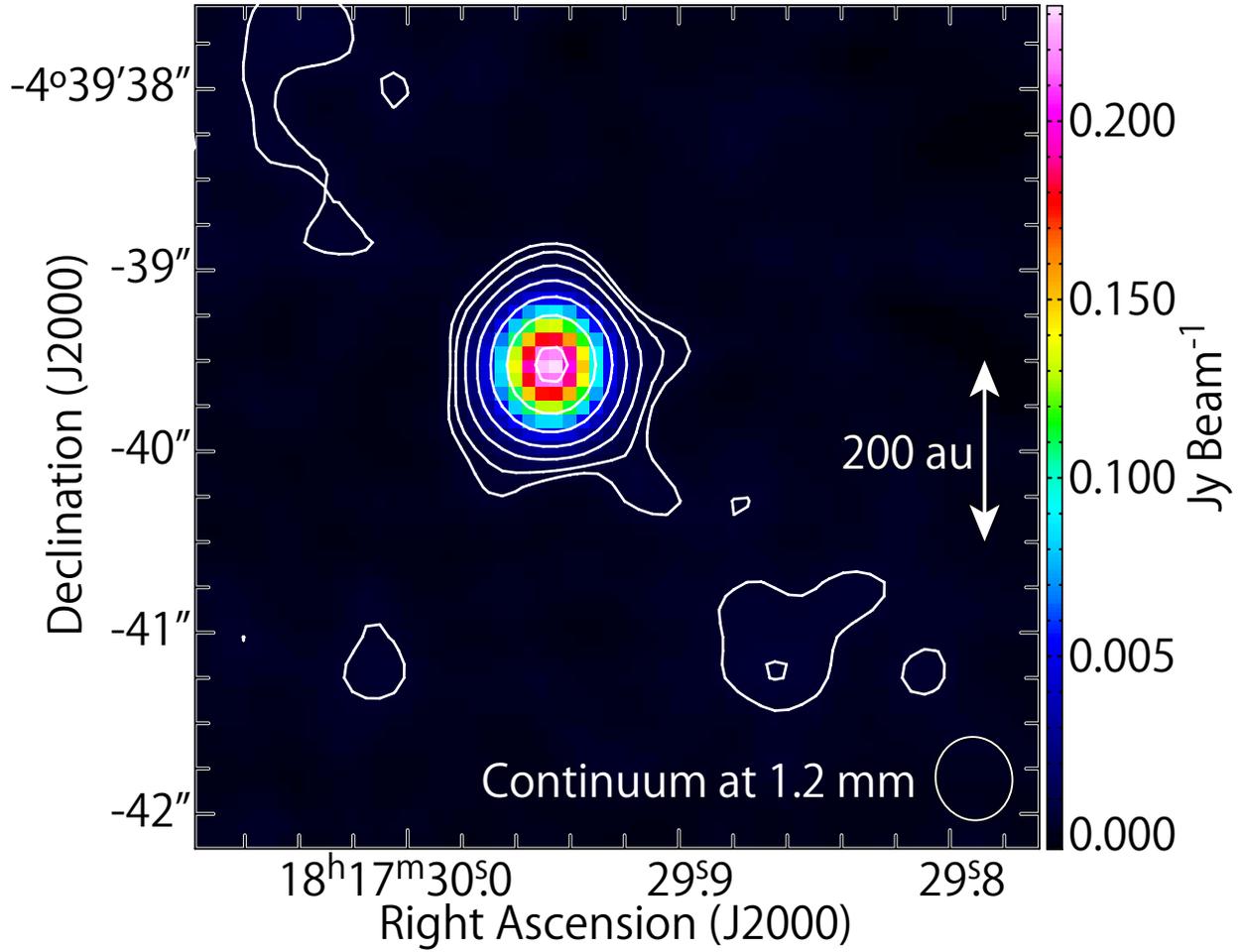}
	\vspace*{-20pt}
	\caption{A map of the dust continuum at 1.2 mm. 
			The contour levels are 3, 5, 10, 20, 40, 80, and 160$\sigma$, 
			where the rms level is 0.13 mJy beam\inv. 
			The synthesized beam size is $0\farcs46 \times 0\farcs42$ (P.A. 11.\!\!\degr76). 
			\label{fig:continuum}}
	\end{center}
\end{figure}

Figure \ref{fig:continuum} shows the map of the 1.2 mm dust continuum, 
where the synthesized beam size is $0\farcs46 \times 0\farcs42$ (P.A. 11.\!\!\degr76). 
Its peak position is determined by the two dimensional Gaussian fit to be: 
($\alpha_{2000}$, $\delta_{2000}$) = (18$^{\rm h}$17$^{\rm m}$29\fs947, -04\degr39$^\prime$39\farcs55). 
The deconvolved size of the continuum emission is $0\farcs23 \times 0\farcs16$ (P.A. 158\degr), 
and hence, the image is not resolved. 
The total flux of the continuum is 28 mJy. 

\begin{landscape}
\begin{figure}
	\begin{center}
	\epsscale{1.0}
	\plotone{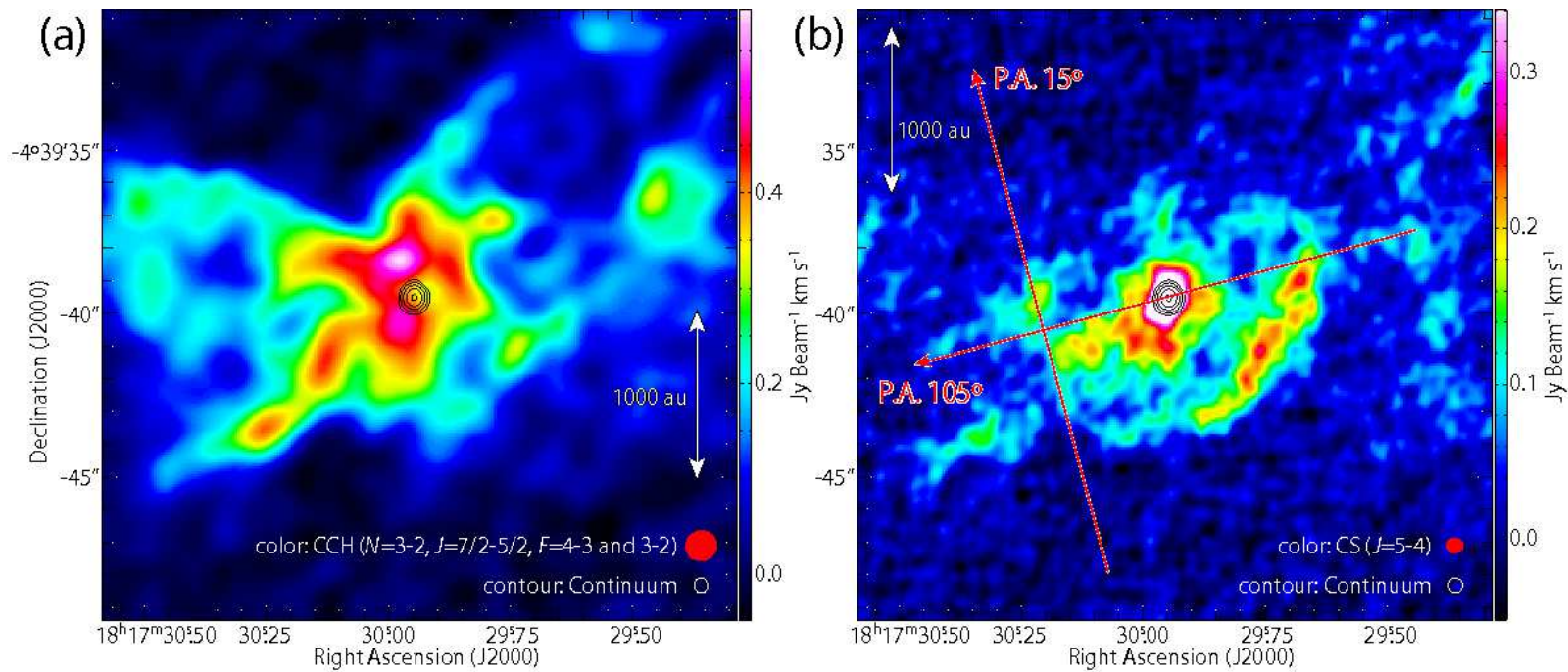}
	\vspace*{-20pt}
	\caption{Integrated intensity maps of CCH (\cch; a) and CS (\cs; b). 
			The black contours represent the 1.2 mm continuum map, 
			where the contour levels are 10, 20, 40, 80, and 160$\sigma$, where the rms level is 0.13 mJy beam\inv. 
			The outflow axis is along the red arrow with a P.A. of 105\degr\ in panel (b). 
			The PV diagram in Figure \ref{fig:PV_CSoutflow} is prepared along the red arrow with a P.A. of 15\degr\ in panel (b), 
			which is centered at the position with an offset of 4\arcsec\ to the southeast from the continuum peak along a P.A. of 105\degr. 
			\label{fig:mom0_extended}}
	\end{center}
\end{figure}
\end{landscape}

\begin{landscape}
\begin{figure}
	\begin{center}
	\epsscale{1.0}
	\plotone{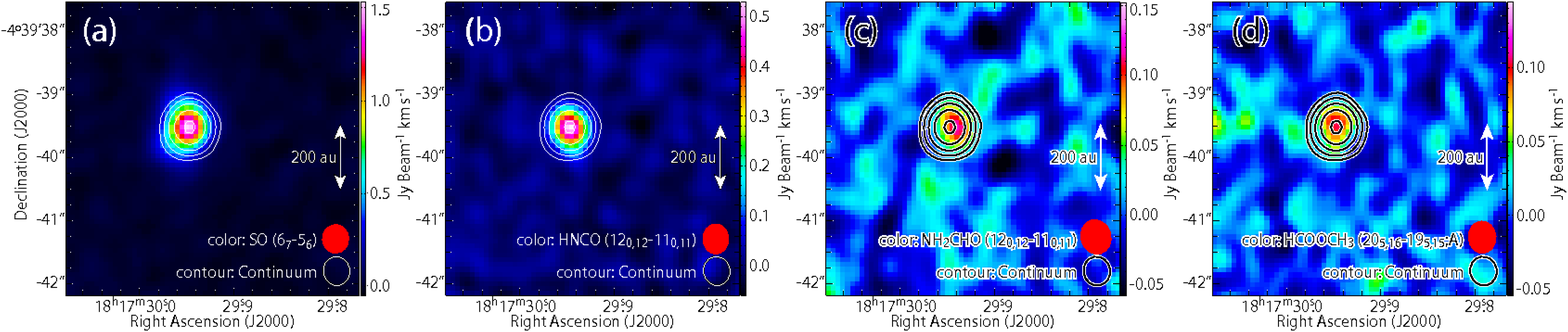}
	\vspace*{-20pt}
	\caption{Integrated intensity maps of SO (\so; a), HNCO (\hnco; b), \FA\ (\fa; c), and \MF\ (\mf; d). 
			The contours represent the 1.2 mm continuum map, 
			where the contour levels are the same as those in Figure \ref{fig:mom0_extended}. 
			\label{fig:mom0_compact}}
	\end{center}
\end{figure}
\end{landscape}

\begin{figure}
	\begin{center}
	\epsscale{1.0}
	\plotone{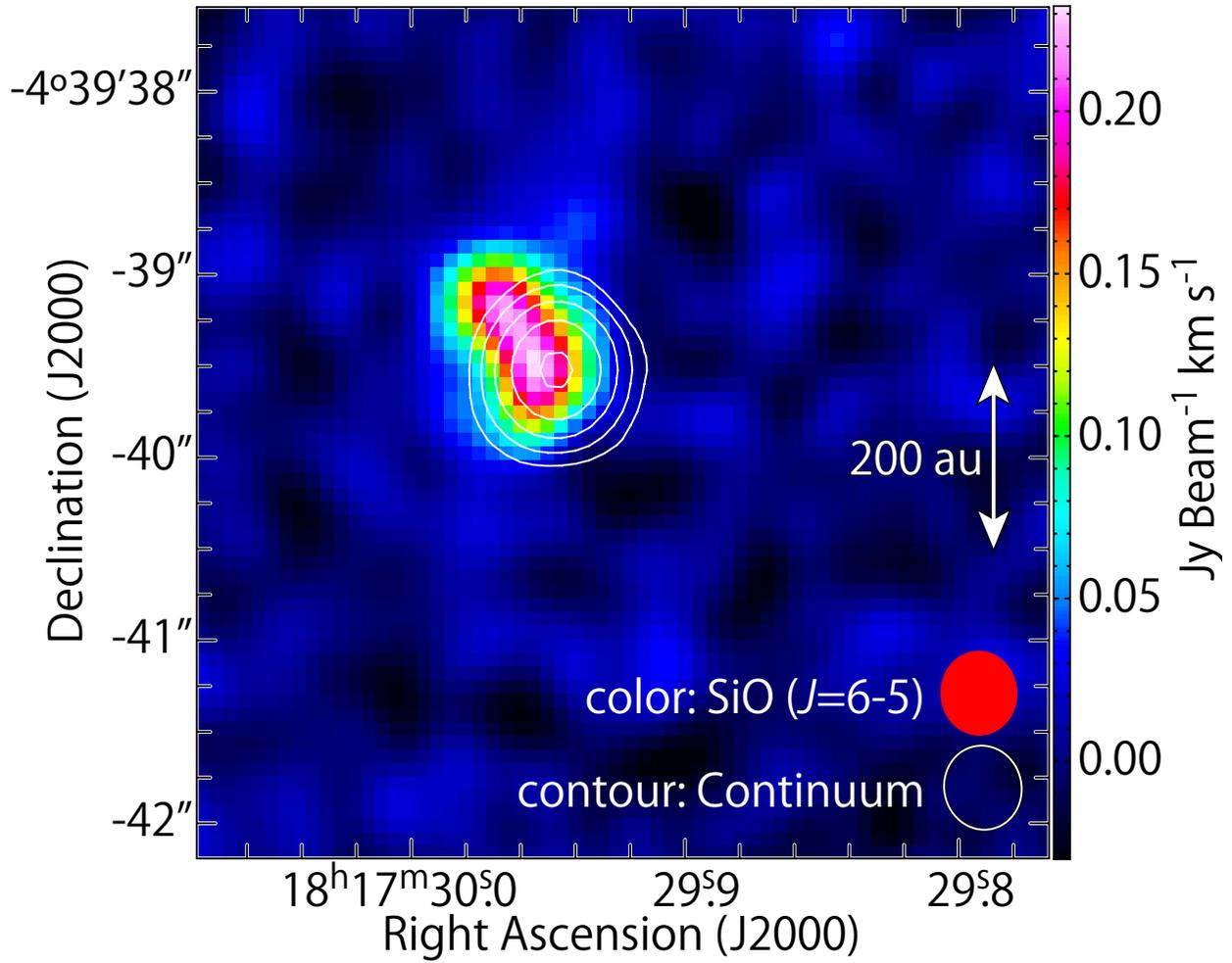}
	\vspace*{-20pt}
	\caption{An integrated intensity map of SiO (\sio). 
			The white contours represent the 1.2 mm continuum map, 
			where the contour levels are the same as those in Figure \ref{fig:mom0_extended}. 
			\label{fig:mom0_SiO}}
	\end{center}
\end{figure}

In this observation, we detected the lines of CCH, CS, SO, HNCO, \FA, \MF, and SiO.  
This observation toward L483 is a part of a larger project to delineate physical and chemical structures of 
the disk forming regions in several protostellar sources with ALMA (\#2013.1.01102.S; P.I. N. Sakai), 
and the above lines are all detected in the other sources (B335 and NGC 1333 IRAS 4A) 
with the same frequency setup (Imai et al. 2016; A. L{\'o}pez-Sepulcre et al. in preparation). 
Hence, their detections are secure, although only one line was detected for each of these species except for CCH. 
As mentioned in Section \ref{sec:intro}, this source is regarded as a WCCC candidate source (Sakai et al. 2009a; Hirota et al. 2009). 
Hence, detections of COMs such as \FA\ and \MF, 
which are characteristic of hot corinos, are notable (e.g. Sakai \& Yamamoto 2013). 
The integrated intensity maps of CCH, CS, SO, HNCO, \FA, \MF, and SiO are 
shown in Figures \ref{fig:mom0_extended}--\ref{fig:mom0_SiO}. 

The CCH (\cch) emission is extended over a 10\arcsec\ scale, 
and hence, the outertaper of 1\arcsec\ is applied to improve a signal-to-noise ratio of the image (Figure \ref{fig:mom0_extended}a). 
The existence of the carbon-chain molecule CCH around the protostar at a few 100 au scale confirms the WCCC nature of this source. 
The CCH distribution has a hole with a radius of $\sim$0\farcs5 around the continuum peak. 
This feature is similar to that found in the WCCC sources L1527 and IRAS 15398--3359 (Sakai et al. 2014a, 2014b; Oya et al. 2014), 
which would originate from the gas-phase destruction and/or depletion onto dust grains. 
The hole of the distribution seems to have a slight offset from the continuum peak to the western side, 
which implies an asymmetric distribution of the gas in the vicinity of the protostar.  
This asymmetry may be related to inhomogeneities of the initial gas distribution. 
In addition to the envelope component, a part of the CCH emission seems to trace an outflow cavity wall. 
The direction of the outflow axis looks consistent with the previous reports 
(P.A. 95--105\degr) (e.g. Park et al. 2000; Tafalla et al. 2000; Chapman et al. 2013). 

The CS (\cs) emission also traces the component extended over a 10\arcsec\ (Figure \ref{fig:mom0_extended}b). 
In addition, it shows a compact component concentrated to the continuum peak. 
The deconvolved size of this compact component is $1\farcs26\times0\farcs88$, 
and is slightly more extended than the 1.2 mm dust continuum. 
This slightly extended component will be discussed in Section \ref{sec:vel_env}. 
On the other hand, the SO (\so), HNCO (\hnco), \FA\ (\fa), and \MF\ (\mf) distributions are highly concentrated to the continuum peak 
(Figure \ref{fig:mom0_compact}). 
The sizes of the distributions deconvolved by the synthesized beam are 
$0\farcs56\times0\farcs39$, and $0\farcs26\times0\farcs16$ for SO and HNCO, respectively. 
The HNCO distribution is almost point-like. 
Similarly, the distributions of \FA\ and \MF\ are 
also point-like at the resolution of the observations. 
In addition to \FA\ and \MF, we tentatively detected the t-HCOOH emission concentrated near the protostar. 

The distribution of SiO (\sio) is different from those of the above molecular species. 
It has a slight extension toward the northeastern direction from the continuum peak, as shown in Figure \ref{fig:mom0_SiO}. 
The extension is significant, considering the synthesized beam of this observation ($\sim 0\farcs5$). 
The size of the SiO distribution deconvolved by the synthesized beam is $0\farcs70\times0\farcs43$ (P.A. 35\degr). 

\section{Velocity Structure} \label{sec:vel_env}
\subsection{Geometrical Configuration of the Disk/Envelope System and the Outflow} \label{sec:env_rotation}
\begin{landscape}
\begin{figure}
	\begin{center}
	\epsscale{1.0}
	\plotone{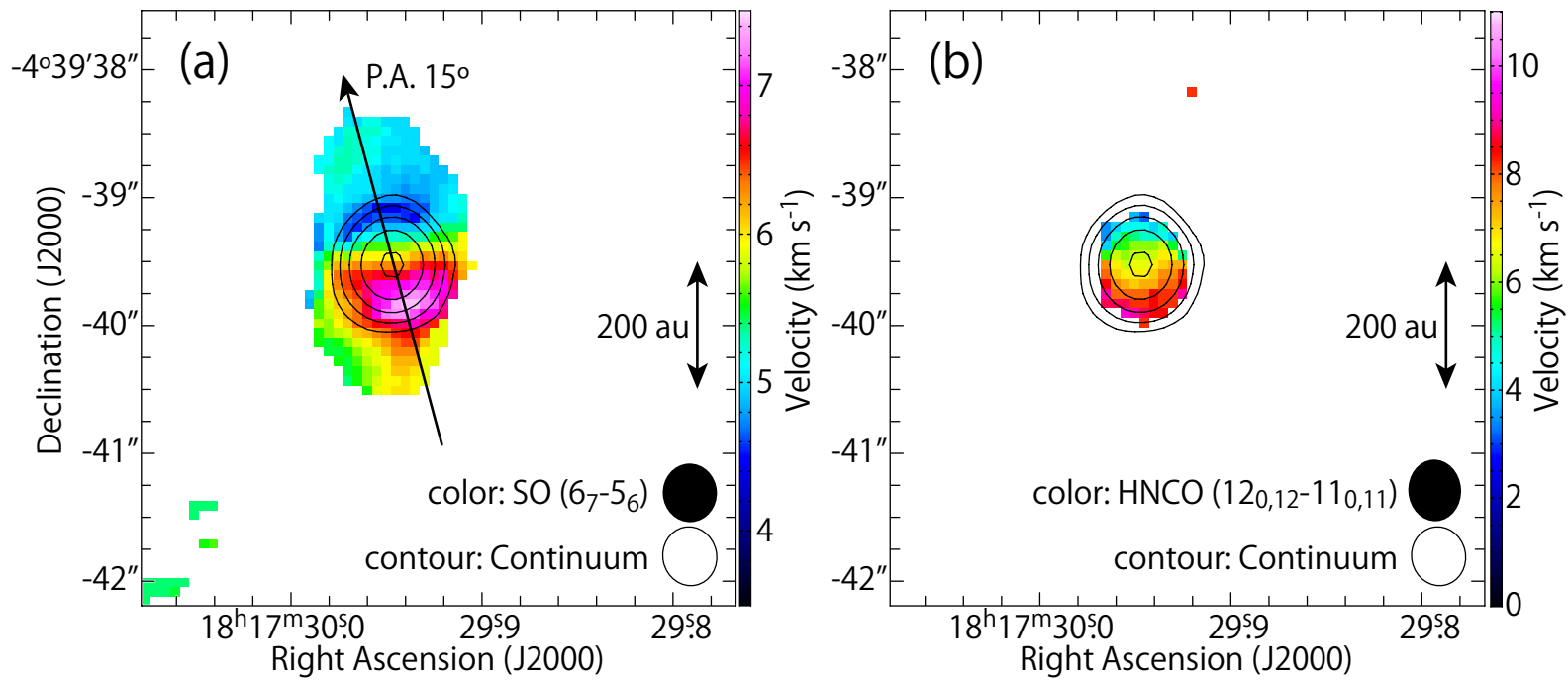}
	\vspace*{-20pt}
	\caption{Moment 1 maps of SO (\so; a) and HNCO (\hnco; b). 
			The black contours represent the 1.2 continuum map, 
			where the contour levels are the same as those in Figure \ref{fig:mom0_extended}. 
			The PV diagrams in Figures \ref{fig:PV_CS_envPA15deg}, 
			\ref{fig:PV_SO-HNCO_envPA15deg}, 
			\ref{fig:PV_CS_IRE-Keplermodel}, and \ref{fig:PV_SO_Keplermodel}  
			are prepared along the black arrow in panel (a) (P.A. 15\degr) and along the direction perpendicular to it 
			centered at the continuum peak. 
			\label{fig:mom1}}
	\end{center}
\end{figure}
\end{landscape}

\begin{figure}
	\begin{center}
	\epsscale{0.8}
	\plotone{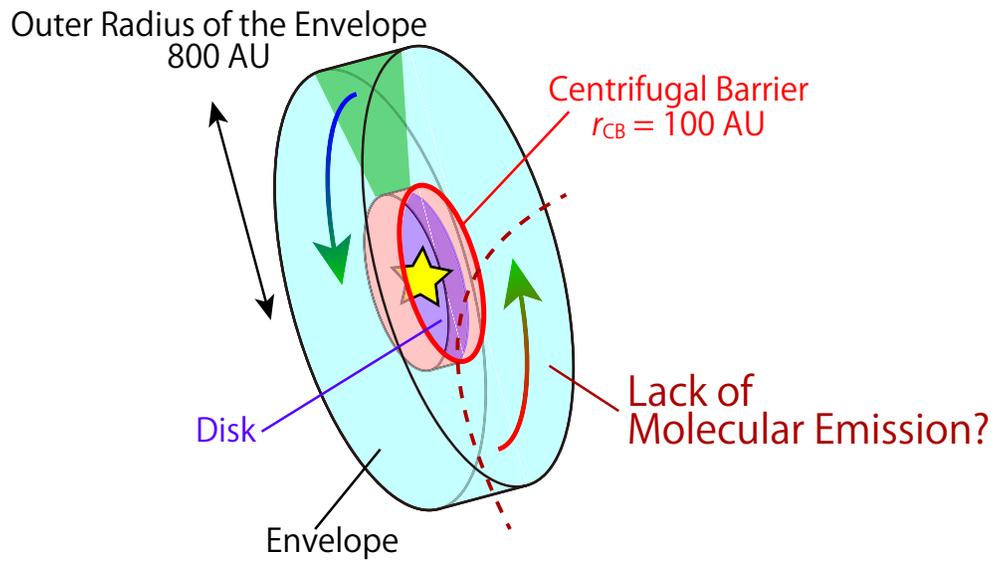}
	\vspace*{-20pt}
	\caption{Schematic illustration of the disk/envelope system in L483. 
			The mid-plane of the disk/envelope is extended along a P.A. of 15\degr, 
			and its western side faces the observer. 
			The line emission in the western side of the protostar seems to be missing, 
			according to the observation (see text). 
			\label{fig:geometry}}
	\end{center}
\end{figure}

Here, we focus on the kinematic structure of the gas concentrated around the protostar. 
In the moment 1 maps of SO and HNCO (Figure \ref{fig:mom1}), 
we see a velocity gradient perpendicular to the outflow axis 
(P.A. 95--105\degr) (e.g. Park et al. 2000; Tafalla et al. 2000; Chapman et al. 2013), 
which strongly suggests rotation. 
A component in the northern side of the continuum peak is blue-shifted, 
while a component in the southern side is red-shifted. 
The direction of the gradient is qualitatively consistent with the CS ($J$=7--6) and HCN ($J$=4--3) observations 
reported by Takakuwa et al. (2007) at a 20\arcsec\ scale. 
In this paper, we employ the position angle of the outflow axis of 105\degr\ reported by Chapman et al. (2013), 
and assume that the disk/envelope system is extended along the position angle of 15\degr. 
Since the eastern and western lobes of the outflow are reported to be red-shifted and blue-shifted, respectively 
(e.g. Fuller et al. 1995; Hatchell et al. 1999; Park et al. 2000; Tafalla et al. 2000; J\o rgensen 2004; Takakuwa et al. 2007; Leung et al. 2016), 
the northwestern side of the disk/envelope system is thought to face the observer, as illustrated in Figure \ref{fig:geometry}. 
In the following subsections, 
the kinematic structures of the observed molecular distributions are described, 
where CCH is excluded from this analysis due to heavy blending of the hyperfine components and a poor S/N ratio. 

\subsection{CS} \label{sec:env_CS}
\begin{figure}
	\begin{center}
	\epsscale{1.0}
	\plotone{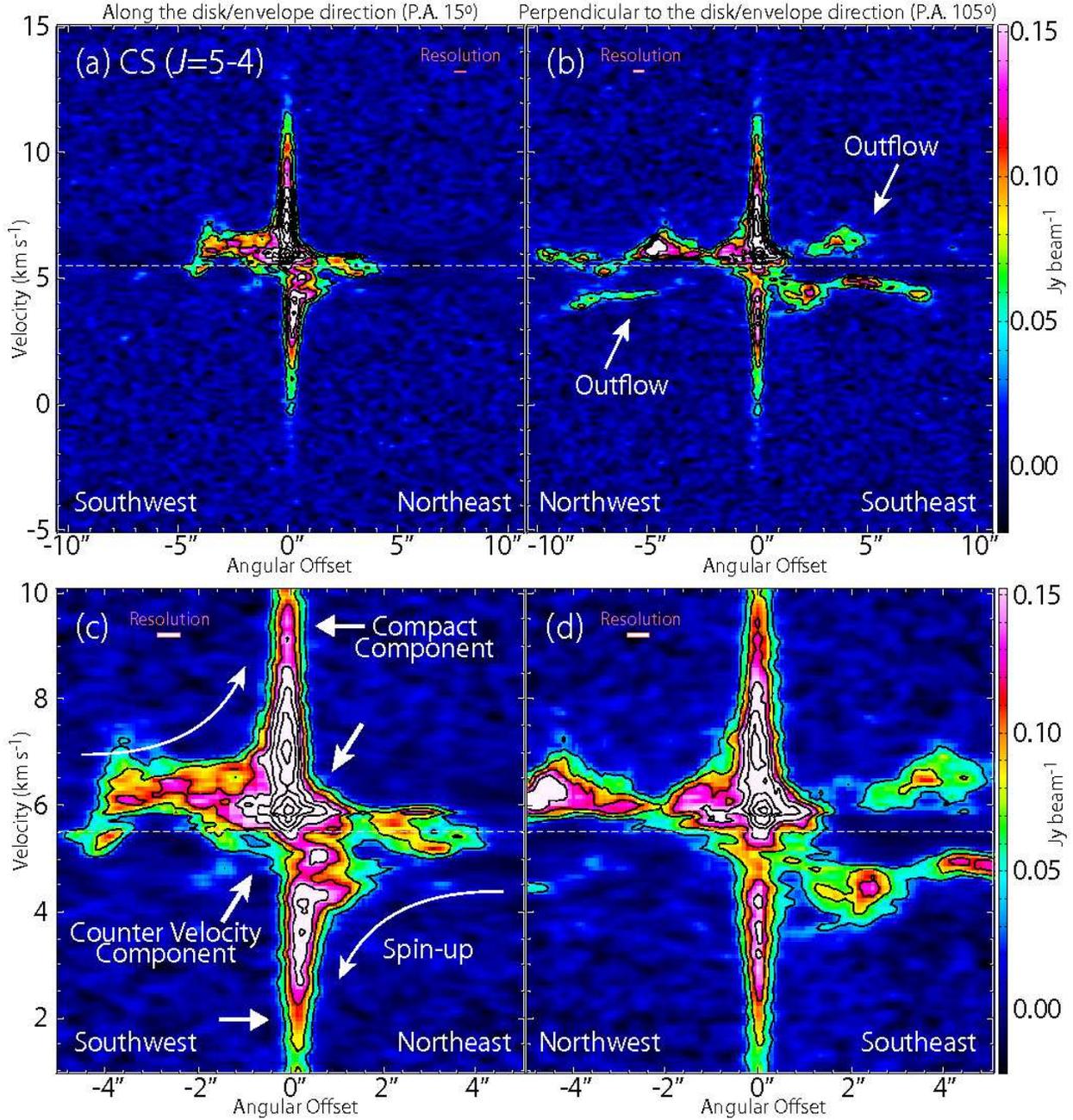}
	\vspace*{-20pt}
	\caption{Position-velocity diagrams of CS (\cs) 
			along the disk/envelope direction (P.A. 15\degr; a, c) indicated in Figure \ref{fig:mom1}, 
			and the direction perpendicular to it (P.A. 105\degr; b, d). 
			Panels (c, d) are the blow-ups of the central parts of panels (a, c). 
			The contour levels are every 5$\sigma$, where the rms level is 7.6 mJy beam\inv. 
			The white dashed lines represent the systemic velocity (5.5 km s\inv). 
			The rectangle in the top-left corner of each panel represents the spatial and velocity resolutions. 
			\label{fig:PV_CS_envPA15deg}}
	\end{center}
\end{figure}

Figure \ref{fig:PV_CS_envPA15deg} shows the position-velocity (PV) diagrams of CS (\cs) along the disk/envelope direction (P.A. 15\degr). 
It shows a spin-up feature to the protostar along the disk/envelope direction at an 8\arcsec\ scale (Figures \ref{fig:PV_CS_envPA15deg}a, c). 
Namely, the southwestern and northeastern sides of the protostar are red-shifted and blue-shifted, respectively. 
At the same time, a weak feature of the counter velocity component, 
which is blue-shifted at the southwestern side and red-shifted at the northeastern side, 
can be seen, although the component at the southwestern side is marginal. 
This feature is specific to the infalling-rotating envelope, 
which is most clearly demonstrated in L1527 (Sakai et al. 2014a). 
Hence, CS likely exists in the infalling-rotating envelope. 
In addition, CS shows a compact high velocity-shift component concentrated toward the protostar, 
whose maximum velocity shift is as high as about 6 km s\inv. 

\begin{figure}
	\begin{center}
	\epsscale{1.0}
	\plotone{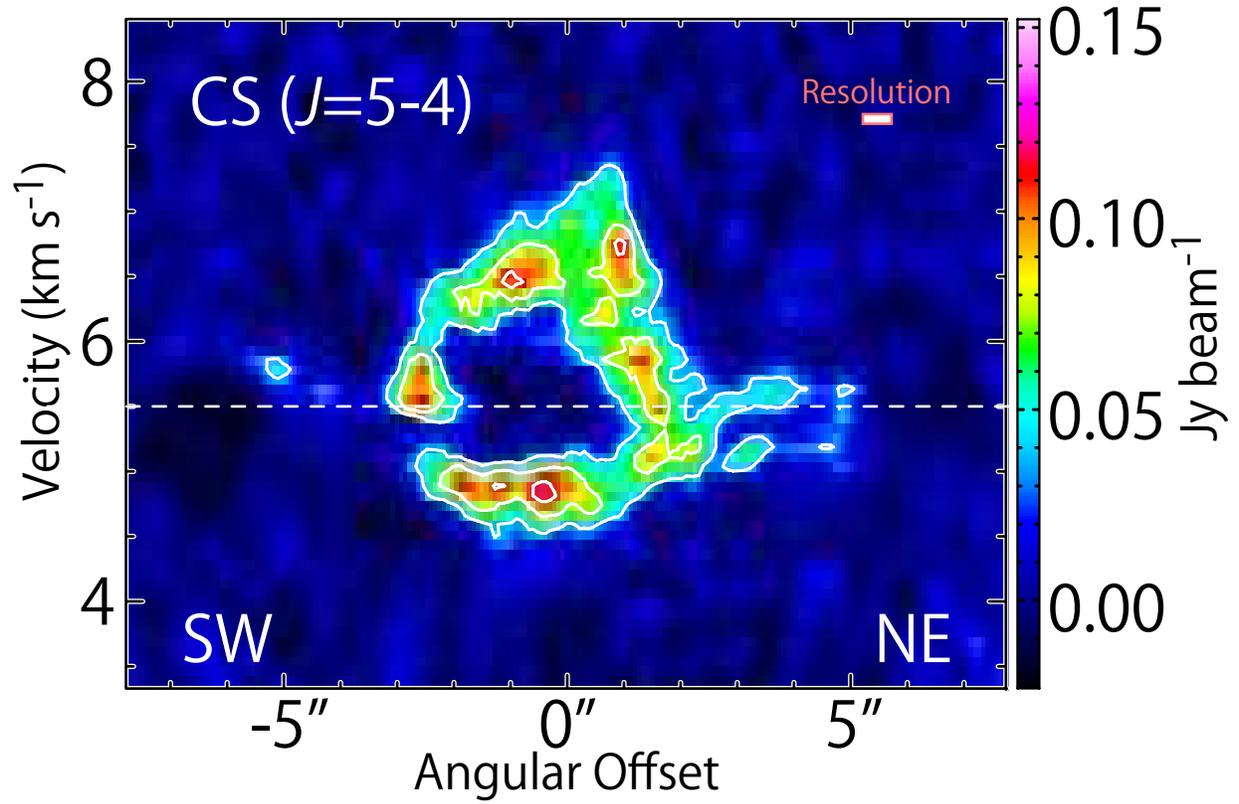}
	\vspace*{-20pt}
	\caption{Position-velocity diagrams of CS (\cs) along the line across the outflow indicated in Figure \ref{fig:mom0_extended}(b). 
			The contour levels are every 5$\sigma$, where the rms level is 7.6 mJy beam\inv. 
			\label{fig:PV_CSoutflow}}
	\end{center}
\end{figure}

Along the line perpendicular to the disk/envelope direction (P.A. 105\degr), 
a component extended over a 20\arcsec\ scale is observed (Figure \ref{fig:PV_CS_envPA15deg}b). 
Although this component looks complicated in the velocity structure, 
it likely traces a part of the outflow cavity. 
Figure \ref{fig:PV_CSoutflow} shows the PV diagram along the line across the outflow lobe in the southeastern side of the protostar, 
which is indicated by an arrow in Figure \ref{fig:mom0_extended}(b). 
The elliptic feature of the PV diagram characteristic of the outflow cavity wall is clearly observed. 
Although both the red-shifted and blue-shifted components with respect to the sytemic velocity (5.5 km s\inv; Hirota et al. 2009) are seen, 
the center velocity of the elliptic feature is slightly red-shifted. 
This elliptic feature is consistent with the configuration illustrated in Figure \ref{fig:geometry}. 
It is quite similar to that observed in the nearly edge-on outflow system of IRAS 15398--3359 (Oya et al. 2014), 
where both the red-shifted and blue-shifted components can be seen in each outflow lobe. 
Hence, it is most likely that the outflow of L483 
blows nearly on the plane of the sky at least in the vicinity of the protostar in contrast to the previous reports (e.g. Fuller et al. 1995). 
Hence, the disk/envelope system likely 
has a nearly edge-on geometry ($i \sim$ 80\degr). 
The detailed structure of the outflow will be reported in a separate publication (Y. Oya et al. in preparation). 

Assuming that the outflow axis is perpendicular to the mid-plane of the disk/envelope system, 
the northwestern side of the disk/envelope system will face the observer. 
If there is an infall motion in the envelope component, 
the southeastern and northwestern sides of the protostar would be blue-shifted and red-shifted, respectively, 
in the PV diagram along the P.A. of 105\degr\ (Figure \ref{fig:PV_CS_envPA15deg}d). 
However, this feature is not clearly recognized, 
because of overwhelming contributions from the outflow component and the missing of the blue-shifted component. 
The velocity gradient due to the infalling motion will be verified with the aid of the kinematic model in Section \ref{sec:IREmodel}. 
On the other hand, the high velocity-shift component does not show any velocity gradient 
along the P.A. of 105\degr\ around the protostar position.

\subsection{SO and HNCO} \label{sec:env_SO-HNCO}
\begin{figure}
	\begin{center}
	\epsscale{1.0}
	\plotone{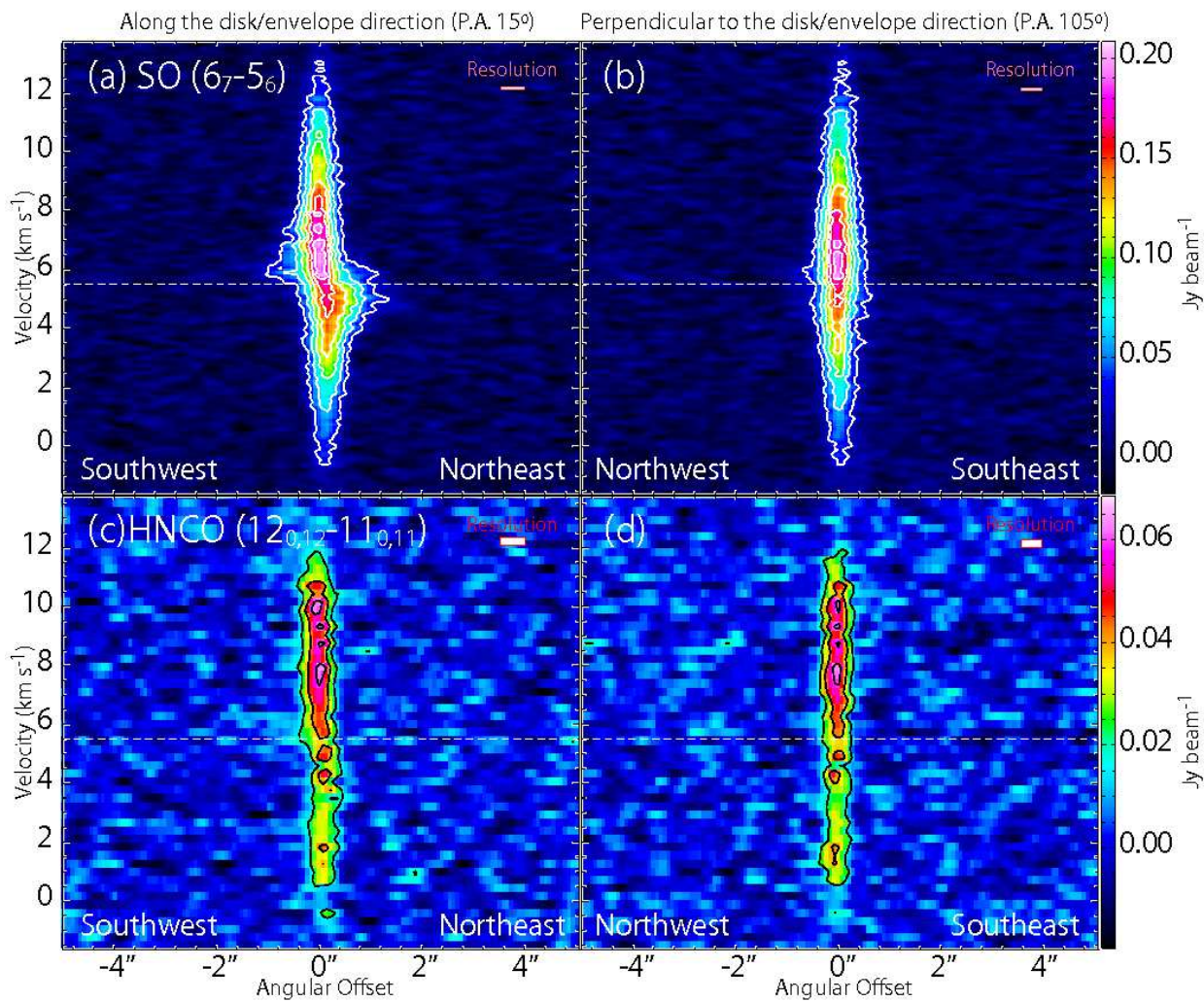}
	\vspace*{-20pt}
	\caption{Position-velocity diagrams of SO (\so; a, b) and HNCO (\hnco; c, d)  
			along the disk/envelope direction (P.A. 15\degr; a, c) indicated in Figure \ref{fig:mom1}, 
			and the direction perpendicular to it (P.A. 105\degr; b, d). 
			The contour levels are every 5 and 3$\sigma$, where the rms levels are 6.1 and 6.5 mJy beam\inv, 
			for SO and HNCO, respectively. 
			\label{fig:PV_SO-HNCO_envPA15deg}}
	\end{center}
\end{figure}

The PV diagrams of SO (\so) and HNCO (\hnco) are shown in Figure \ref{fig:PV_SO-HNCO_envPA15deg}. 
Along the disk/envelope direction, a slight velocity gradient is seen for SO (Figure \ref{fig:PV_SO-HNCO_envPA15deg}a), 
while it is scarcely recognized for HNCO (Figure \ref{fig:PV_SO-HNCO_envPA15deg}c). 
This velocity gradient in SO is consistent with that found in CS, 
and hence, it seems to originate from the rotation motion in the inner part of the disk/envelope system. 
No definitive velocity gradient along the line perpendicular to the disk/envelope direction 
can be seen for SO and HNCO in Figures \ref{fig:PV_SO-HNCO_envPA15deg}(b, d). 
Hence, the SO and HNCO emission reveals no significant infalling motion in the vicinity of the protostar. 

\begin{figure}
	\begin{center}
	\epsscale{0.3}
	\plotone{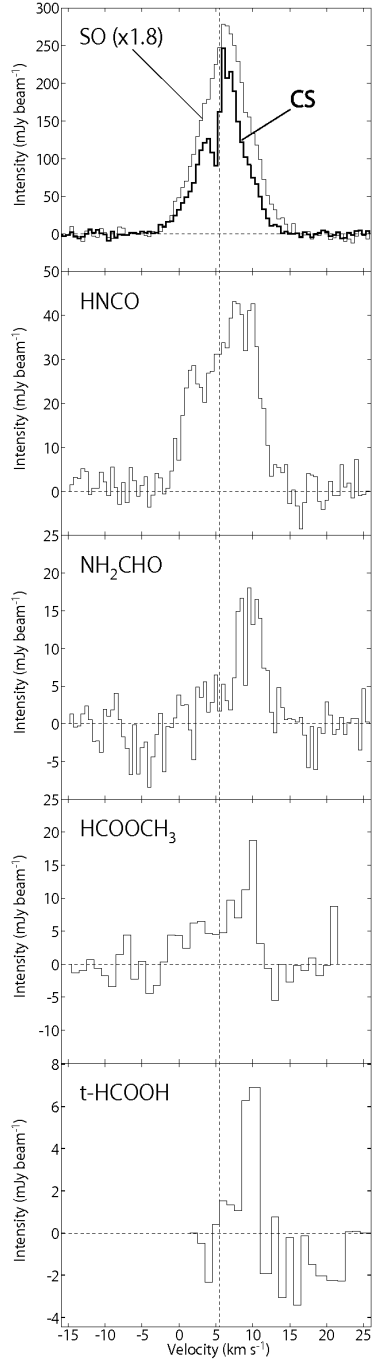}
	\vspace*{-20pt}
	\caption{Spectral line profiles of CS (\cs), SO (\so), HNCO (\hnco), \FA\ (\fa), \MF\ (\mf), and t-HCOOH (\hcooh) toward the protostar position. 
			The line intensities are averaged in a circular region with a diameter of 0\farcs5 centered at the continuum peak. 
			The original spectra are smoothed to improve the signal-to-noise ratio, 
			so as that the velocity resolution is 
			0.5 \kmps\ for CS, SO, HNCO, and \FA, and 1 \kmps\ for \MF\ and t-HCOOH. 
			\label{fig:spectra}}
	\end{center}
\end{figure}

The high velocity-shift components of SO and HNCO near the protostar position 
correspond to that found in the CS emission (Figure \ref{fig:PV_CS_envPA15deg}). 
Figure \ref{fig:spectra} shows the line profiles of CS, SO, and HNCO 
in a circular region with a diameter of 0\farcs5 centered at the continuum peak. 
The line profile of SO shows a profile similar to that of CS, except for the self-absorption in CS. 
Their broad line widths reflect the high velocity-shift component concentrated around the protostar shown 
in Figures \ref{fig:PV_CS_envPA15deg}, \ref{fig:PV_SO-HNCO_envPA15deg}(a), and \ref{fig:PV_SO-HNCO_envPA15deg}(b). 
HNCO also shows a component whose velocity shift from the systemic velocity is larger than 5 \kmps. 
However, the red-shifted component is brighter than the blue-shifted component. 
This feature is also seen in its PV diagrams (Figures \ref{fig:PV_SO-HNCO_envPA15deg}c, d). 
This implies that the HNCO distribution is asymmetric in the vicinity of the protostar. 
This asymmetry implies a less amount of gas in the southwestern part of the disk/envelope system (Figure \ref{fig:geometry}). 

\subsection{\FA\ and \MF} \label{sec:env_COMs}
The most notable result in this study is detections of the COMs, \FA\ and \MF. 
These two species are concentrated around the protostar, as shown in the moment 0 maps (Figures \ref{fig:mom0_compact}c, d). 
The spectral line profiles of \FA\ and \MF\ toward the protostar position 
are dominated by a red-shifted component (Figure \ref{fig:spectra}), 
which is similar to the HNCO spectrum. 
Since the red-shifted component is enhanced for all the HNCO, \FA, and \MF\ spectra consistently, their detections are secure; 
i.e. they do not correspond to other species/transitions. 
The red-shifted line profile implies an asymmetric distribution of those molecules in the vicinity of the protostar, as mentioned in Section \ref{sec:env_SO-HNCO}. 
However, an origin of the asymmetry is puzzling, and is left for future high spatial resolution observations. 

\section{Analysis with the Infalling-Rotating Envelope Model} \label{sec:IREmodel}
In order to understand the chemical differentiation observed above in terms of the physical structure around the protostar, 
we analyze the kinematic structure of the disk/envelope system. 
In L1527, TMC--1A, and IRAS 16293--2422 A, 
the kinematic structures of the infalling-rotating envelopes are successfully explained by a simple ballistic model 
(Sakai et al. 2014a, 2014b, 2016; Oya et al. 2015, 2016). 
Hence, we apply the same model to the CS data. 
However, we cannot reproduce the overall velocity structure with the infalling-rotating envelope model. 
As explained below, we need to consider the two physical components: 
the infalling-rotating envelope and the centrally concentrated component. 

\begin{figure}
	\begin{center}
	\epsscale{1.0}
	\plotone{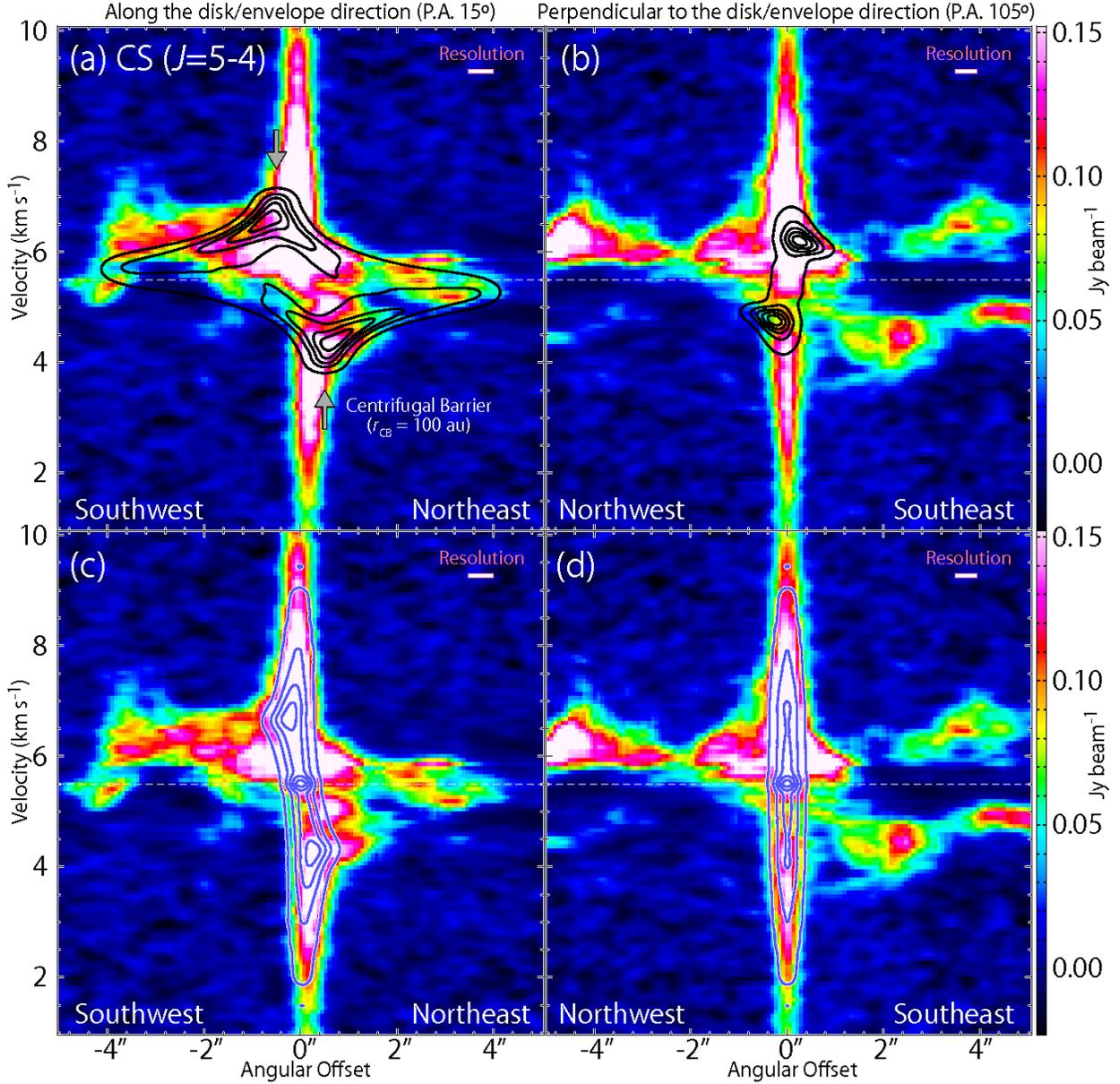} 
	\vspace*{-20pt}
	\caption{Position-velocity diagrams of CS (\cs), 
			where the color maps are the same as the panels (c, d) in Figure \ref{fig:PV_CS_envPA15deg}. 
			The black contours in panels (a, b) represent the results of the infalling-rotating envelope models, 
			where $M$ = 0.15 \Msun, \rCB\ = 100 au, and $i$ = 80\degr. 
			The blue contours in panels (c, d) represent the results of the Keplerian model with the above $M$ and $i$ values, 
			where the emission is simply assumed to come from the inside of the centrifugal barrier. 
			In panels (a--d), the intrinsic line width is assumed to be 0.2 \kmps, 
			and the model image is convolved with the synthesized beam. 
			The contour levels are every 20\% from 5\% of each peak intensity. 
			\label{fig:PV_CS_IRE-Keplermodel}}
	\end{center}
\end{figure}

The infalling-rotating envelope model employed in this study is essentially the same as that reported by Oya et al. (2014). 
Since we are interested in the velocity structure of the infalling-rotating envelope, 
the key model parameters are 
the protostellar mass ($M$), 
the radius of the centrifugal barrier (\rCB), and the inclination angle of the disk/envelope system ($i$).  
Here, we assume the flattened envelope with a constant thickness (30 au), 
which has the radial density distribution of $r^{-1.5}$, for simplicity. 
Unfortunately, the key parameters can loosely be constrained from these observations 
because of the contamination of the overwhelming centrally-concentrated component. 
Hence, we calculate the models with various parameters to find the reasonable set of the parameters by eye. 
Figures \ref{fig:PV_CS_IRE-Keplermodel}a and \ref{fig:PV_CS_IRE-Keplermodel}b shows an example of the simulation of the infalling-rotating envelope which reproduces the observed PV diagrams 
as much as possible except for the central high-velocity components. 
The model parameters are the protostellar mass ($M$) of 0.15 \Msun, 
the radius of the centrifugal barrier (\rCB) of 100 au. 
Since a nearly edge-on geometry is suggested by the outflow structure (Section \ref{sec:env_CS}), 
we roughly assume an inclination angle respect to the plane of the sky ($i$) of 80\degr\ (0\degr\ for a face-on configuration) 
(Y. Oya et al. in preparation). 
The infalling motion along the direction perpendicular to the disk/envelope system can marginally be recognized 
in the PV diagram with the aid of the model (Figure \ref{fig:PV_CS_IRE-Keplermodel}b). 
Its blue-shifted part is missing probably due to the asymmetric gas distribution mentioned above (Figure \ref{fig:geometry}). 

\begin{figure}
	\begin{center}
	\epsscale{1.0}
	\plotone{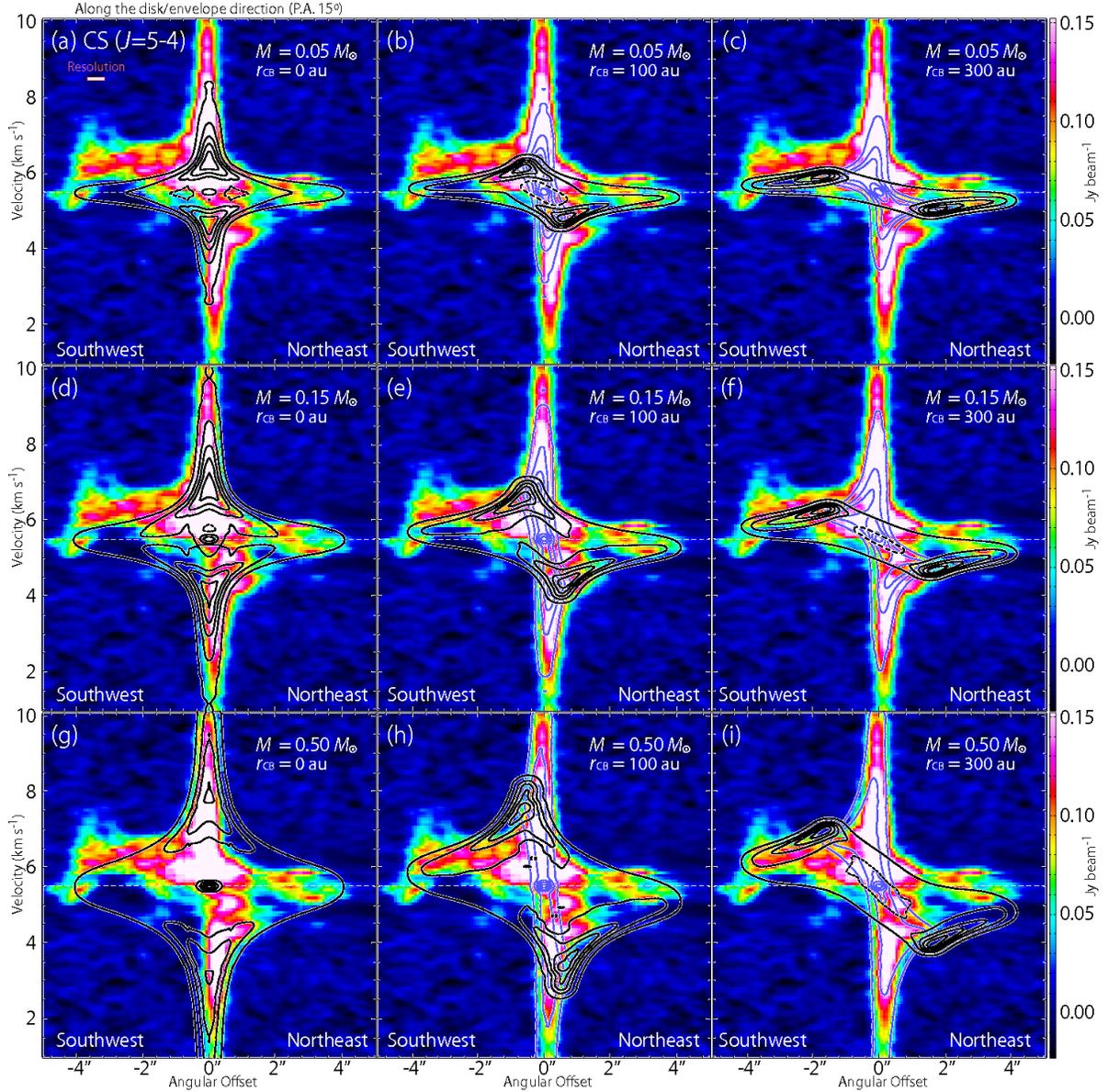}
	\vspace*{-20pt}
	\caption{Position-velocity diagram of CS (\cs) along the disk/envelope direction (P.A. 15\degr), 
			where the color maps are the same as the panel (c) in Figure \ref{fig:PV_CS_envPA15deg}. 
			The black contours represent the results of the infalling-rotating envelope models. 
			The parameters are: $M$ = 0.05, 0.15, and 0.5 \Msun; \rCB\ = 0, 100, and 300 au; $i$ = 80\degr. 
			The blue contours represent the results of the Keplerian model 
			with the same physical parameters as those for the infalling-rotating envelope model in each panel. 
			In the Keplerian model, the emission is simply assumed to come from the inside of the centrifugal barrier. 
			The contour levels are every 20\% from 5\% of the peak intensity of each model. 
			The dashed contours around the central position in the (a, b, d, f, h, i) panels represent the dip toward the center. 
			\label{fig:PV_CS_various_PA015deg}}
	\end{center}
\end{figure}

\begin{figure}
	\begin{center}
	\epsscale{1.0}
	\plotone{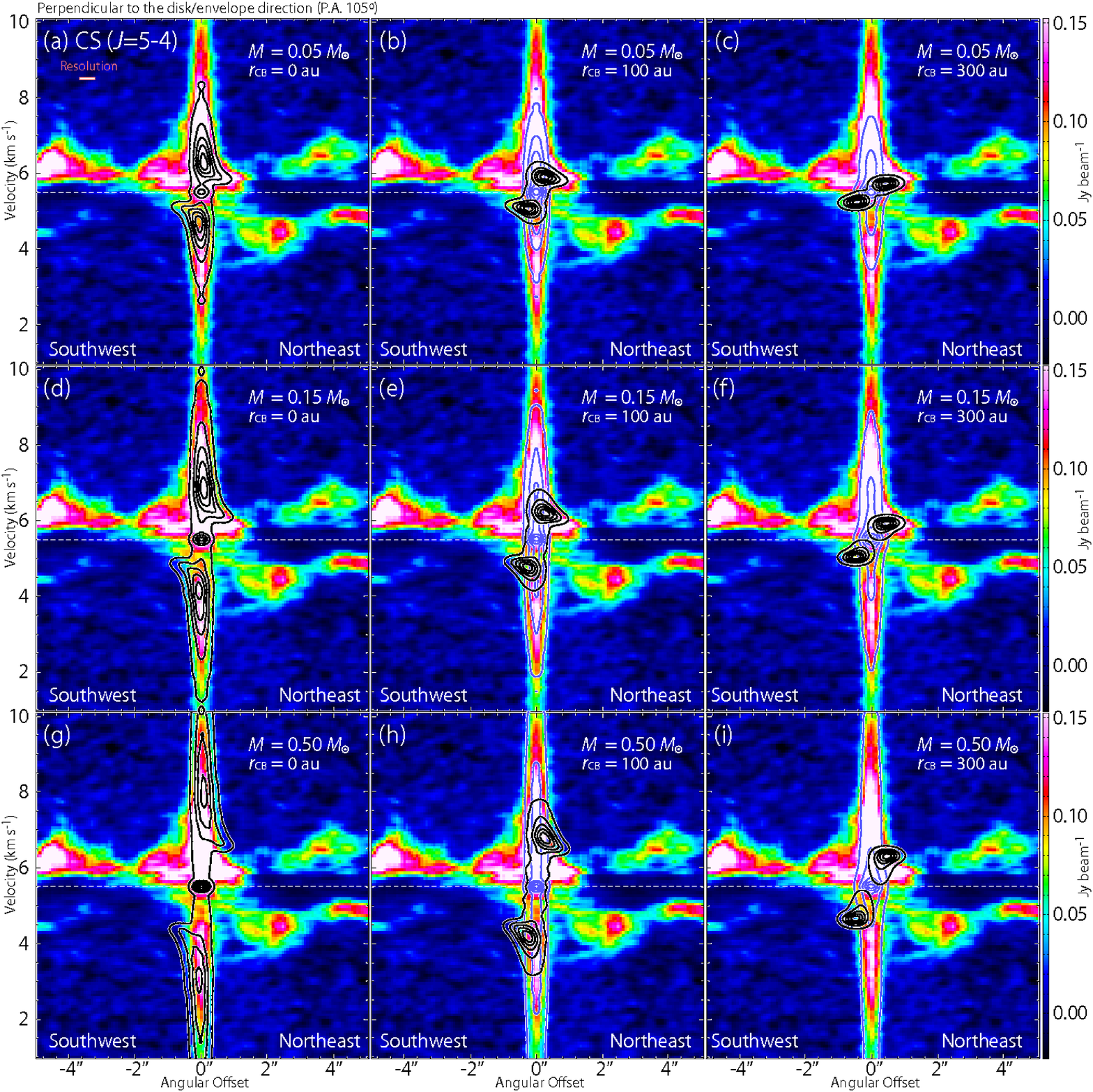}
	\vspace*{-20pt}
	\caption{Position-velocity diagram of CS (\cs) along the direction perpendicular to the disk/envelope direction (P.A. 105\degr), 
			where the color maps are the same as the panel (d) in Figure \ref{fig:PV_CS_envPA15deg}. 
			The black contours represent the results of the infalling-rotating envelope models. 
			The parameters are: $M$ = 0.05, 0.15, and 0.5 \Msun; \rCB\ = 0, 100, and 300 au; $i$ = 80\degr. 
			The blue contours represent the results of the Keplerian model 
			with the same physical parameters as those for the infalling-rotating envelope model in each panel. 
			In the Keplerian model, the emission is simply assumed to come from the inside of the centrifugal barrier. 
			The contour levels are every 20\% from 5\% of the peak intensity of each model. 
			\label{fig:PV_CS_various_PA105deg}}
	\end{center}
\end{figure}

To see how the model PV diagram depends on the radius of the centrifugal barrier (\rCB) and the protostellar mass ($M$), 
we also conducted the simulations of the PV diagrams along the disk/envelope direction and along the direction perpendicular to it {\bf by using the infalling-rotating envelope model with} various sets of these two parameters, 
as shown in Figures \ref{fig:PV_CS_various_PA015deg} and \ref{fig:PV_CS_various_PA105deg}, respectively.  
For the case of no rotation (\rCB\ = 0 au), 
the positions of the most blue-shifted component and the most red-shifted component coincide in the model, 
which apparently contradict with the observation (Figure \ref{fig:PV_CS_various_PA015deg}). 
On the other hand, the velocity of the counter velocity component, 
which represents the infalling motion of the infalling-rotating envelope, 
is underestimated for the \rCB\ = 300 au case. 
As for the mass of the protostar, 0.05 \Msun\ and 0.5 \Msun\ do not reproduce the PV diagram. 
Above all, a reasonable agreement is obtained, 
except for the central high-velocity components, 
for the range of \rCB\ from 30 au to 200 au and the range of the protostellar mass from 0.1 \Msun\ to 0.2 \Msun. 
Hence, the \rCB\ of 100 au and the protostellar mass of 0.15 \Msun\ are chosen as the representative values, as mentioned above (Figure \ref{fig:PV_CS_IRE-Keplermodel}). 
For more stringent constraints, further detailed analysis with a high angular resolution observation is needed. 
The infalling motion has a higher velocity shift than the observation 
if the model parameters are set to explain the high velocity-shift component centrally concentrated near the protostar. 
This justifies the two component model consisting of the infalling-rotating envelope and the centrally concentrated component described above. 
Although we can see some excess red-shifted emission in the southwestern part of the PV diagram (Figure \ref{fig:PV_CS_IRE-Keplermodel}a), 
the model can roughly reproduce the infalling-rotating envelope part of the PV diagrams. 
We analyze the high velocity component in a separate way, as described below. 

\begin{figure}
	\begin{center}
	\epsscale{1.0}
	\plotone{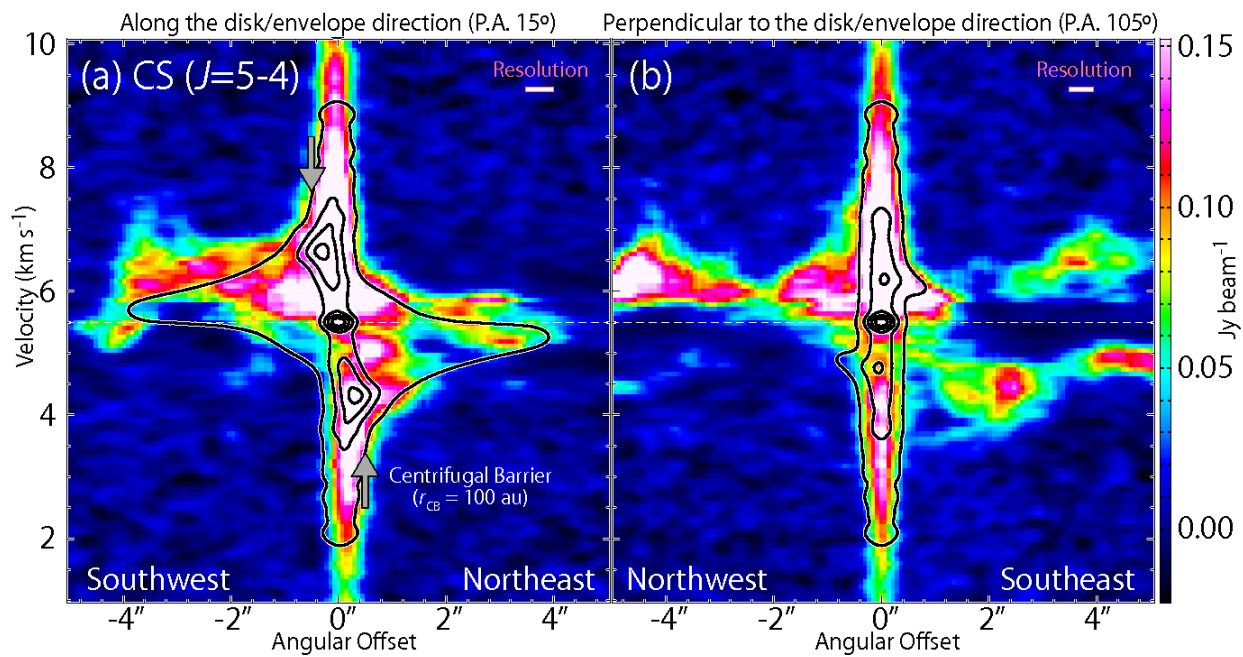}
	\vspace*{-20pt}
	\caption{Position-velocity diagrams of CS (\cs) 
			along the disk/envelope direction (P.A. 15\degr; a) 
			and the direction perpendicular to it (P.A. 105\degr; b), 
				where the color maps are the same as the panels (c, d) in Figure \ref{fig:PV_CS_envPA15deg}. 
				The black contours represent the results of an infalling-rotating envelope model combined with the Keplerian model, 
				where the physical parameters are the same as those for the models for Figure \ref{fig:PV_CS_IRE-Keplermodel}. 
				The contour levels are every 20\% from 3\% of each peak intensity. 
				\label{fig:PV_CS_IRE-Kepler-merged}}
	\end{center}
\end{figure}

\begin{figure}
	\begin{center}
	\epsscale{1.0}
	\plotone{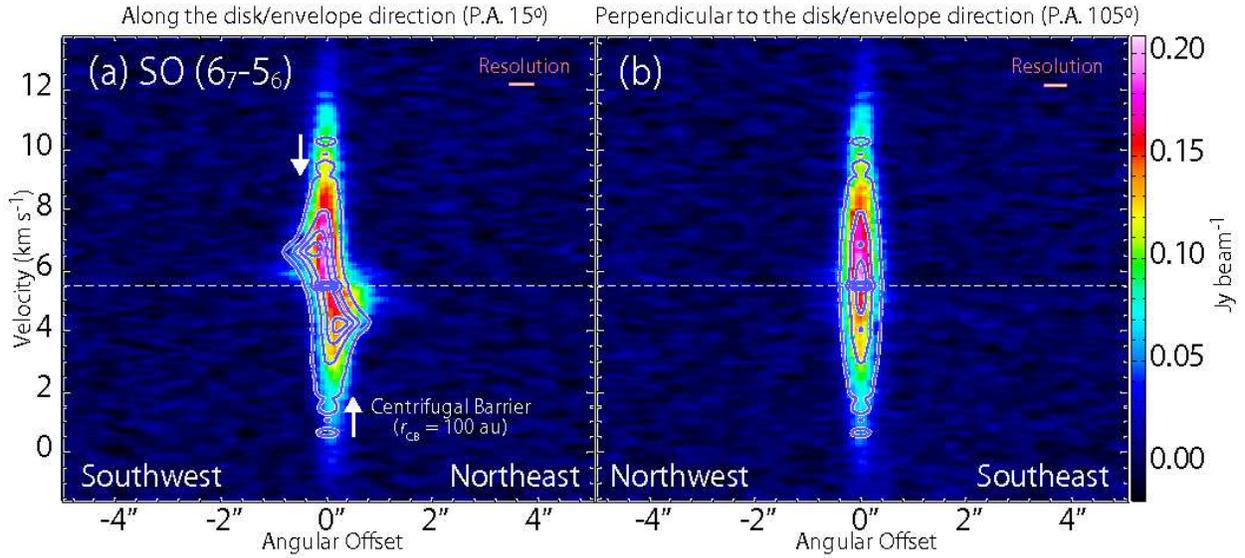}
	\vspace*{-20pt}
	\caption{Position-velocity diagrams of SO (\so), 
			where the color maps are the same as those in Figures \ref{fig:PV_SO-HNCO_envPA15deg}(a, b). 
			The blue contours 
			represent the results of the Keplerian model, 
			where $M$ = 0.15 \Msun\ and $i$ = 80\degr, 
			and the emission is simply assumed to be only inside the centrifugal barrier (100 au). 
			The contour levels are every 20\% from 3\% of each peak intensity. 
			\label{fig:PV_SO_Keplermodel}}
	\end{center}
\end{figure}

The most likely candidate for the centrally-concentrated high-velocity component traced by CS, SO, HNCO, \FA, and \MF\ 
is the Keplerian disk inside the centrifugal barrier, although the rotation curve is not resolved. 
Assuming $M$ of 0.15 \Msun\ and $i$ of 80\degr,  
which are roughly estimated from the above analysis of the infalling-rotating envelope, 
we can reproduce the high velocity-shift part of the PV diagrams of CS and SO by the Keplerian disk model 
(Figures \ref{fig:PV_CS_IRE-Keplermodel} and \ref{fig:PV_SO_Keplermodel}). 
A radius of the emitting region of the maximum velocity component ($\sim$6 \kmps) in the disk is estimated to be as small as 4 au. 
We also show the results of the Keplerian disk model combined with the infalling-rotating envelope model in Figure \ref{fig:PV_CS_IRE-Kepler-merged}. 
In addition, the results of the Keplerian disk model are overlaid on those of the infalling-rotating envelope model in Figures \ref{fig:PV_CS_various_PA015deg} and \ref{fig:PV_CS_various_PA105deg} for reference. 

In L483, the kinematic structure of the gas traced by CS around the protostar is similar to those of \FAD\ in L1527 and \HHCS\ in IRAS 16293--2422 A, 
which are explained by a combination of the infalling-rotating envelope component 
and the (possible) Keplerian disk component inside the centrifugal barrier (Sakai et al. 2014b; Oya et al. 2016). 
Hence, the CS distribution in L483 is different from those in L1527 and TMC--1A (Sakai et al. 2014b, 2016), 
where CS resides only in the infalling-rotating envelope. 
Such a distribution of CS in L483 seems to resemble the IRAS 16293--2422 A case. 
In IRAS 16293--2422 A, the emission of the normal isotopic species of CS seems to come from the disk 
as well as the infalling-rotating envelope (Y. Oya et al. in preparation), 
although the C$^{34}$S emission traces the envelope outside the centrifugal barrier (Favre et al. 2014). 
The difference of the behavior of CS would originate from the higher bolometric luminosity of 
L483 (13 \Lsun; Shirley et al. 2000) and IRAS 16293--2422 A (22 \Lsun; Crimier et al. 2010) 
than L1527 (1.7 \Lsun; Green et al. 2013) and TMC--1A (2.5 \Lsun; Green et al. 2013). 
The higher bolometric luminosity would cause the higher temperature of the disk component inside the centrifugal barrier, 
which prevents the CS depletion in this region. 
Since the binding energy of CS is 1900 K 
(UMIST Database for Astrochemistry; McElroy et al. 2013, http://udfa.ajmarkwick.net/index.php), 
the evaporation temperature is about 40 K \citep[]{Yamamoto_chembook}. 
This is just above the mid-plane temperature of the disk just inside the centrifugal barrier in L1527 (30 K; Sakai et al. 2014b). 
If the midplane temperature inside the centrifugal barrier is higher in L483 due to the higher bolometric luminosity, 
CS does not freeze out. 

On the other hand, the rotation feature of the SO emission 
revealed in the PV diagrams (Figure \ref{fig:PV_SO-HNCO_envPA15deg}a) 
looks similar to that in L1527 (Sakai et al. 2014a), 
where SO mainly highlights the centrifugal barrier. 
However, the high velocity-shift components concentrated toward the protostar are much brighter in L483 than in L1527, 
and hence, the rotation motion traced by SO in L483 is expected to come mainly from the disk component inside the centrifugal barrier (Figure \ref{fig:PV_SO_Keplermodel}). 
The relatively high bolometric luminosity in L483 could again help in escaping SO from depletion onto dust grains in the disk component. 

Assuming that the SO emission appears inside of the centrifugal barrier, 
we can directly estimate its radius. 
The doconvolved size (FWHM) is 0\farcs5 (100 au) along the disk/envelope direction (P.A. 15\degr), 
and hence, the radius of the centrifugal barrier is estimated to be $\sim$50 au. 
Since this size will be affected by the strong emission from the vicinity of the protostar, 
it can be regarded as the lower limit. 
On the other hand, 
the 5 $\sigma$ contour in the PV diagram of SO (Figure \ref{fig:PV_SO-HNCO_envPA15deg}a) 
is extended to $2\farcs4$ (= 480 au), 
where the radius deconvolved with the beam size along the position angle of 15\degr\ is $\sim$1\arcsec\ ($\sim$ 200 au). 
Since the SO line may trace the envelope component just outside the centrifugal barrier (Sakai et al. 2017), 
this size can be regarded as the upper limit for the radius of the centrifugal barrier. 
Although it is difficult to derive the radius of the centrifugal barrier from the SO emission 
because of the contamination by the disk component in L483, 
these sizes will directly give rough estimates for the size of the centrifugal barrier. 
They are consistent with the estimate from the analysis of the PV diagrams of CS. 

\section{SiO Emission} \label{sec:outflow_SiO}
\begin{figure}
	\begin{center}
	\epsscale{1.0}
	\plotone{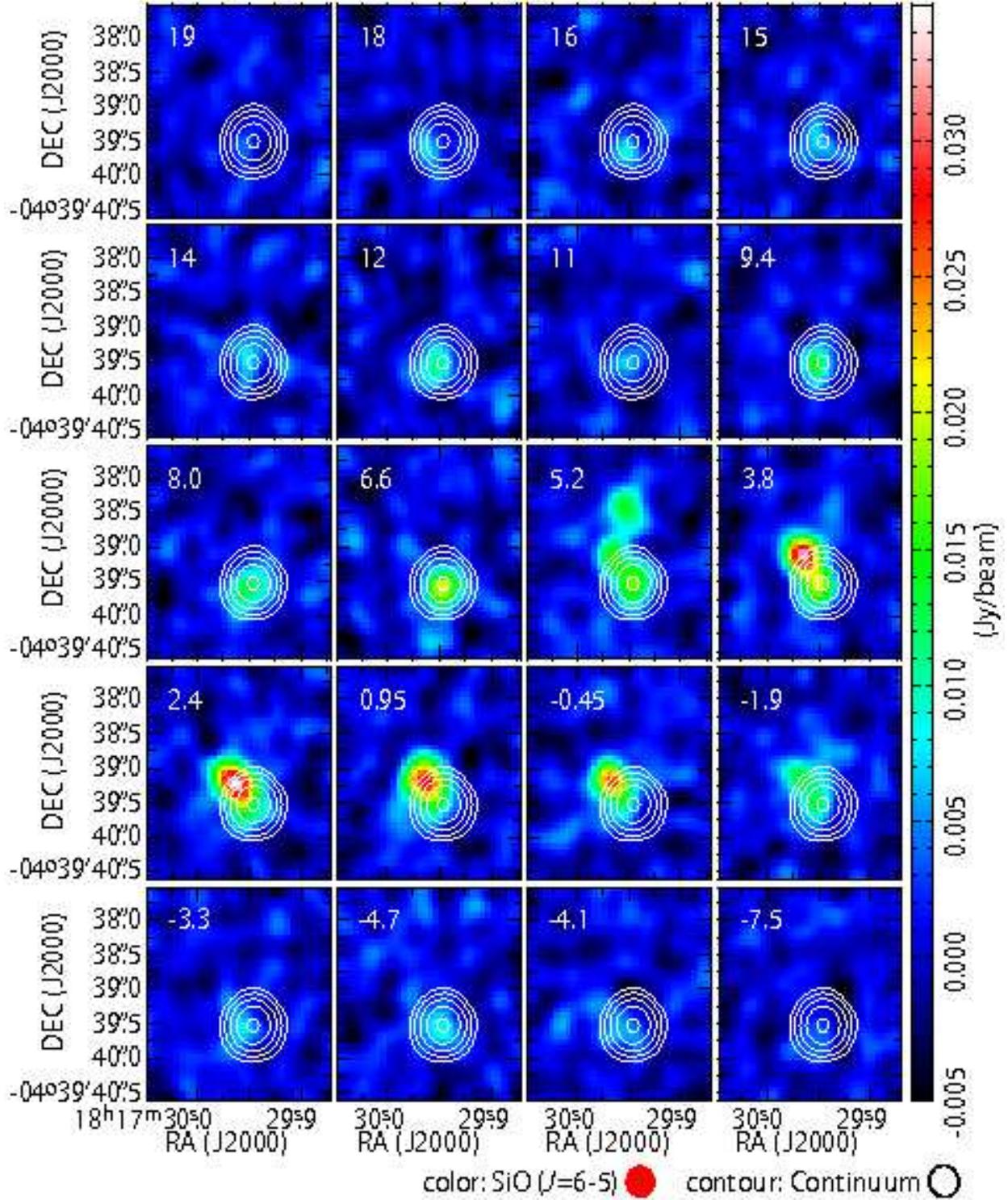}
	\vspace*{-30pt}
	\caption{Velocity channel maps of SiO (\sio; color). 
			Each map represents the averaged intensity with the velocity width of 1.4 km s\inv. 
			The white contours represent the 1.2 mm continuum map, 
			where the contour levels are the same as those in Figure \ref{fig:mom0_extended}. 
			The value in the top left corner of each panel is the averaged velocity (km s\inv). 
			\label{fig:channelmap_SiO}}
	\end{center}
\end{figure}

The distribution of SiO extends to the northeastern direction from the continuum peak, 
as mentioned in Section \ref{sec:distribution}. 
The position angle of the direction of this extension is $\sim$35\degr\ (Figure \ref{fig:mom0_SiO}). 
Figure \ref{fig:channelmap_SiO} shows the velocity channel maps of SiO. 
The blue-shifted component of SiO (\vlsr\ $<$ 3.8 km s\inv) is offset from the protostar by 0\farcs5 ($\sim$ 100 au), 
while the weak red-shifted component appears at the protostar position. 
The position of the blue-shifted component of SiO is close to the expected position of the centrifugal barrier (\rCB\ = 100 au) or inside of it. 
Since its velocity is faster than that of CS, 
it does not seem to come from a part of the infalling-rotating envelope. 
It is most likely that this component represents the shock caused by the outflow. 
The existence of such a shocked gas near the centrifugal barrier is puzzling. 
It might be related to the launching mechanism of the outflow, 
although the association of the shocked gas with the centrifugal barrier has to be explored at a higher angular resolution. 

\section{Chemical Composition} \label{sec:chem}
L483 is proposed to have the WCCC character on the basis of the single-dish observations 
(Sakai et al. 2009a; Hirota et al. 2010; Sakai \& Yamamoto 2013). 
It is confirmed by detection of CCH 
in the vicinity of the protostar at a 1000 au scale. 
Nevertheless, \FA\ and \MF, which are related to hot corino chemistry, are also detected. 
Furthermore, we tentatively detected the weak lines of t-HCOOH and \AAL. 
This is the first spatially-resolved detection of saturated COMs in the WCCC source. 
Although such a mixed chemical character source has been recognized as an intermediate source in previous studies (Sakai et al. 2009a), 
the present observation confirms its definitive existence of such a source at a high angular resolution. 

The beam averaged column densities of \FA\ and \MF\ toward the protostar position are evaluated 
by assuming local thermodynamic equilibrium at 70, 100, and 130 K (Table \ref{tb:abundance}). 
This range of temperatures is typical of hot corinos (e.g. Oya et al. 2016). 
The beam-averaged column densities of \FA\ and \MF\ are calculated to be 
$(1.5 \pm 0.7) \times 10^{14}$ cm$^{-2}$ and $(7 \pm 4) \times 10^{15}$ cm$^{-2}$, respectively, at 100 K. 
The column densities change by 30\% and 14\% for the change in the assumed temperature by $\pm30$ K 
for \FA\ and \MF, respectively. 
The column density of the tentatively detected species (t-HCOOH) and the upper limits to the column densities of \AAL\ and \DEE\ are also calculated, 
as shown in Table \ref{tb:abundance}. 

To derive the fractional abundances relative to H$_2$, 
the beam-averaged H$_2$ column density is derived from the dust continuum to be $6.5 \times 10^{23}$ cm$^{-2}$ 
by using the following relation (Ward-Thompson et al. 2000): 
\begin{align}
	N ({\rm H}_2) &= \frac{2 \ln 2 \cdot c^2}{\pi h \kappa_\nu m} 
		\times \frac{F(\nu)}{\nu^3 \theta_{\rm major} \theta_{\rm minor}} 
		\times \left(\exp \left(\frac{h\nu}{kT} \right) - 1\right), 
\end{align}
where $M$ is the gas mas, $\kappa_\nu$ is the mass absorption coefficient with respect to the gas mass, 
$m$ is the averaged mass of a particle in the gas ($3.83 \times 10^{-24}$ g), 
$\nu$ is the frequency, $F(\nu)$ is the peak flux, $\theta_{\rm major}$ and $\theta_{\rm minor}$ are the major and minor beam size, respectively, 
$c$ is the speed of light, $h$ is the Planck's constant, 
and $T$ is the dust temperature. 
We here assume the dust temperature of 100 K. 
$\kappa_\nu$ is evaluated to be 0.008 cm$^2$ g\inv\ at 1.2 mm with $\beta$ = 1.8 (Shirley et al. 2011) 
under the assumption that $\kappa_\nu$ depends on the wavelength $\lambda$ 
with the equation: $\kappa_\nu = 0.1 \times \left(0.3 {\rm\ mm} / \lambda \right)^\beta$ cm$^2$ g\inv\ (Beckwith et al. 1990). 
If the dust temperature is 70 K and 130 K, the H$_2$ column density is $9.5 \times 10^{23}$ and $4.9 \times 10^{23}$ cm$^{-2}$, respectively. 
The fractional abundances relative to H$_2$ are then evaluated to be $(1.3-3.9) \times 10^{-10}$ and $(7.3-16.2) \times 10^{-9}$ for \FA\ and \MF, respectively, 
assuming that the gas temperature is the same as the dust temperature (Table \ref{tb:abundance}). 
These fractional abundances of \FA\ and \MF\ are comparable with those reported for the hot corinos, 
IRAS 16293--2422 ($6 \times 10^{-10}$ and $9 \times 10^{-9}$; Jaber et al. 2014) 
and B335 ((1--9) $ \times\ 10^{-10}$ and (2--8) $\times\ 10^{-9}$; Imai et al. 2016). 
L483 indeed harbors a hot corino activity in the closest vicinity of the protostar. 

As mentioned in Section \ref{sec:vel_env}, different molecules trace different parts in this source. 
The infalling-rotating envelope is traced by CS (and possibly CCH), 
while the compact component concentrated in the vicinity of the protostar, 
which is likely to be a disk component inside the centrifugal barrier, 
is traced by CS, SO, HNCO, \FA, and \MF. 
Such a chemical change around the centrifugal barrier is previously reported for some other sources: 
L1527, TMC--1A, and IRAS 16293--2422 A (Sakai et al. 2014b, 2016; Oya et al. 2016). 
However, CS and SO are found to be quite abundant in the compact component concentrated in the vicinity of the protostar in this source 
in contrast to the L1527 and TMC--1A cases. 
The situation similar to L483 is also seen in IRAS 16293--2422 A (Section \ref{sec:env_CS}). 
Thus, the chemical change would be highly dependent on sources. 
Hence, it is still essential to investigate chemical structures of various protostellar sources at a sub-arcsecond resolution. 

\section{Summary} \label{sec:summary}
We observed the Class 0 protostar L483 with ALMA in various molecular lines. 
The major results are as follows: 

\noindent (1)
A chemical differentiation at a 100 au scale is found in L483. 
The CCH emission has a central hole of 0\farcs5 (100 au) in radius. 
The CS emission traces the compact component concentrated near the protostar, 
the extended envelope component, and a part of the outflow cavity. 
In contrast, the SO and HNCO emission only shows the compact component. 

\noindent (2)
In spite of the WCCC character of this source, the saturated COMs, \FA\ and \MF, are detected. 
Their emission is highly concentrated near the protostar. 
This result is the first spatially-resolved example of the mixed character source of WCCC and hot corino chemistry. 

\noindent (3)
The kinematic structures of the envelope and the disk components traced by CS are analyzed 
by simple models of the infalling-rotating envelope and the Keplerian disk in order to understand 
the observed chemical differentiation in terms of the physical structure. 
The protostellar mass of 0.1--0.2 \Msun\ and the radius of the centrifugal barrier of 30--200 au 
roughly explain the infalling-rotating envelope part of the observed PV diagrams of CS. 
The compact component likely traces the Keplerian disk component inside the centrifugal barrier, 
although the rotation curve is not resolved. 
Hence, the above chemical change seems to be occurring around the centrifugal barrier. 

\noindent (4)
In L483, CS, which is thought to be a good tracer of the infalling-rotating envelope, 
traces the disk component as well. 
SO, which highlights the ring structure around the centrifugal barrier in L1527 and TMC--1A, 
also traces the disk component in this source. 
These results would originate from the higher luminosity of L483. 

\noindent (5)
The SiO distribution has an extension from the protostar position. 
It may trace the local outflow-shock near the centrifugal barrier.

\acknowledgments
This paper makes use of the ALMA data set ADS/JAO.ALMA\#2013.1.01102.S. 
ALMA is a partnership of the ESO (representing its member states), 
the NSF (USA) and NINS (Japan), together with the NRC (Canada) and the NSC and ASIAA (Taiwan), 
in cooperation with the Republic of Chile. 
The Joint ALMA Observatory is operated by the ESO, the AUI/NRAO and the NAOJ. 
The authors are grateful to the ALMA staff for their excellent support. 
Y.O. acknowledges the JSPS fellowship. 
This study is supported by Grant-in-Aid from the Ministry of Education, Culture, Sports, Science, and Technologies of Japan (25400223, 25108005, and 15J01610). 
N.S. and S.Y. acknowledge financial support by JSPS and MAEE under the Japan--France integrated action program (SAKURA: 25765VC).
C.C. and B.L. acknowledge the financial support by CNRS under the France--Japan action program.



\clearpage
\begin{landscape}
\begin{table}
	\begin{center}
	\caption{Parameters of the Observed Lines\tablenotemark{a} 
			\label{tb:lines}}
	\vspace*{-20pt} 
	\begin{tabular}{llccccc}
		\hline
		Molecule & Transition & Frequency (GHz) & $E_u$ (K) & $S\mu^2$ (Debye$^2$)\tablenotemark{b} & $A_{ij}$ ($s^{-1}$) & Synthesized Beam \\ \hline
		CS & \cs & 244.9355565 & 35.3 & 19 & 2.98 $\times 10^{-4}$ & $0\farcs51 \times 0\farcs46$ (P.A. -177.\!\!\degr24) \\ 
		\FA\tablenotemark{c} & \fa & 247.390719 & 78.1 & 156 & 1.10 $\times 10^{-3}$ & $0\farcs56 \times 0\farcs49$ (P.A. 16.\!\!\degr44) \\ 
		\MF\tablenotemark{c} & \mf & 249.0474280 & 141.6 & 50 & 1.46 $\times 10^{-3}$ & $0\farcs52 \times 0\farcs45$ (P.A. -177.\!\!\degr18) \\ 
		SiO\tablenotemark{c} & \sio & 260.5180090 & 43.8 & 58 & 9.12 $\times 10^{-4}$ & $0\farcs46 \times 0\farcs42$ (P.A. -177.\!\!\degr78) \\ 
		\AAL\tablenotemark{c} & \aal & 260.5440195 & 96.4 & 82 & 6.25 $\times 10^{-4}$ & $0\farcs46 \times 0\farcs42$ (P.A. -177.\!\!\degr81) \\ 
		SO & \so & 261.8437210 & 47.6 & 16 & 2.28 $\times 10^{-4}$ & $0\farcs46 \times 0\farcs42$ (P.A. 3.\!\!\degr09) \\ 
		CCH\tablenotemark{d} & $N$=3--2, $J$=7/2--5/2,  & 262.0042600 & 25.1 & 2.3 & 5.32 $\times 10^{-5}$ & $0\farcs98 \times 0\farcs92$ (P.A. -78.\!\!\degr06) \\ 
		 & $F$=4--3 and $F$=3--2 & 262.0064820 & & 1.7 & 5.12 $\times 10^{-5}$ & $0\farcs98 \times 0\farcs92$ (P.A. -78.\!\!\degr06) \\ 
		t-HCOOH\tablenotemark{c} & \hcooh & 262.103481 & 82.8 & 24 & 2.03 $\times 10^{-4}$ & $0\farcs46 \times 0\farcs42$ (P.A. 2.\!\!\degr98) \\ 
		\DEE\tablenotemark{c} & \dee & 262.393513 & 118.0 & 148 & 7.18 $\times 10^{-5}$ & $0\farcs46 \times 0\farcs41$ (P.A. 1.\!\!\degr69) \\ 
		HNCO\tablenotemark{c} & \hnco & 263.7486250 & 82.3 & 30 & 2.56 $\times 10^{-4}$ & $0\farcs48 \times 0\farcs41$ (P.A. 3.\!\!\degr87) \\ 
		\hline 
	\end{tabular}
	\tablenotetext{a}{Taken from CDMS (M\"{u}ller et al. 2005) and JPL (Pickett et al. 1998). }
	\tablenotetext{b}{Nuclear spin degeneracy is not included. } 
	\tablenotetext{c}{4 channel binded. The spectral profile of the t-HCOOH line (Figure \ref{fig:spectra}) is obtained with binding 16 channels. } 
	\tablenotetext{d}{An outertaper of 1\arcsec\ is applied. } 
	\end{center}
\end{table}
\end{landscape}

\clearpage
\begin{landscape}
\begin{table}
	\begin{center}
	\caption{Column Densities and Fractional Abundances of the Molecules Observed toward the Protostar Position\tablenotemark{a, b} 
			\label{tb:abundance}}
	\vspace*{10pt}
	\begin{tabular}{lccccccc}
		\hline 
		 & \multicolumn{3}{c}{Column Density ($10^{14}$ cm$^{-2}$)} & \hspace{10pt} & \multicolumn{3}{c}{Fractional Abundance Relative to H$_2$ ($10^{-10}$)} \\ 
		 Dust Temperature & 70 K & 100 K & 130 K && \hspace{30pt}70 K\hspace*{30pt} & 100 K & 130 K \\ \hline
		 CS & $5.1 \pm 0.2$ & $6.2 \pm 0.2$ & $7.5 \pm 0.2$ && $5.3 \pm 0.2$ & $9.6 \pm 0.3$ & $15.2 \pm 0.5$ \\ 
		 HNCO & $9.6 \pm 1.1$ & $11.5 \pm 1.3$ & $14.1 \pm 1.6$ && $10.0 \pm 1.1$ & $17.7 \pm 2.0$ & $28.6 \pm 3.2$ \\ 
		 \FA & $1.2 \pm 0.6$ & $1.5 \pm 0.7$ & $1.9 \pm 0.9$ && $1.3 \pm 0.6$ & $2.4 \pm 1.1$ & $3.9 \pm 1.9$ \\ 
		 \MF & $69.8 \pm 35.9$ & $69.9 \pm 36.0$ & $80.0 \pm 41.2$ && $73 \pm 38$ & $107 \pm 55$ & $162 \pm 84$ \\ 
		 \hdashline \multicolumn{8}{c}{Tentative-detection} \\ 
		 t-HCOOH & $2.0 \pm 1.9$ & $2.4 \pm 2.2$ & $2.9 \pm 2.7$ && $2.1 \pm 1.9$ & $3.6 \pm 3.4$ & $5.8 \pm 5.5$ \\ 
		 \hdashline \multicolumn{8}{c}{Upper Limit} \\ 
		 \AAL & $<0.62$ & $<0.75$ & $<0.95$ && $<0.65$ & $<1.2$ & $<1.9$ \\ 
		 \DEE & $< 116.1$ & $< 119.6$ & $< 135.0$ && $< 121.6$ & $< 183.8$ & $< 273.6$ \\ 
		 \hline
	\end{tabular}
	\tablenotetext{a}{Derived under the assumption of local thermodynamic equilibrium at the temperature of 100K. 
					N(H$_2$) is derived from the 1.2 mm dust continuum emission (See text). } 
	\tablenotetext{b}{Errors are derived from three times the root mean square noise of the integrated intensity. } 
	\end{center}
\end{table}
\end{landscape}

\clearpage

\clearpage

\clearpage

\clearpage

\clearpage

\clearpage

\clearpage

\clearpage

\clearpage

\clearpage

\clearpage

\clearpage

\clearpage

\clearpage

\clearpage

\clearpage

\end{document}